\documentclass[10pt]{article}
\usepackage{amsmath}
\usepackage{amssymb}
\usepackage{graphicx}
\usepackage{cite}
\usepackage{color}
\usepackage{setspace}
\usepackage[labelfont=bf,labelsep=period,justification=raggedright]{caption}

\makeatletter
\renewcommand{\@biblabel}[1]{\quad#1.}
\makeatother

\date{}

\usepackage[numbers]{natbib}
\usepackage{rotating}
\usepackage{booktabs}
\usepackage{colortbl}
\usepackage[table]{xcolor}
\usepackage{float}

\begin{document}

\begin{titlepage}

\begin{center}
{\Large
\textbf{The Dawn of Open Access to Phylogenetic Data}
}
\vfill

Andrew F. Magee$^{1}$, 
Michael R. May$^{1}$, 
Brian R. Moore$^{1,\ast}$

\bigskip

$\mbox{}^1$Department of Evolution and Ecology, University of California, Davis\\
\vspace{-0.4\baselineskip}
Davis, CA 95616, \mbox{U.S.A.} \\

\vfill

R.H. Open Access to Phylogenetic Data
\end{center}

\vfill

\begin{flushleft}
$^{\ast}$Corresponding Author:\\
Brian R. Moore \\
\vspace{-0.4\baselineskip}
Phone: 530-752-7104 \\
\vspace{-0.4\baselineskip}
E-mail: {\tt brianmoore@ucdavis.edu} \\
\end{flushleft}

\end{titlepage}

\newpage
\section*{Abstract}

The scientific enterprise depends critically on the preservation of and open access to published data. 
This basic tenet applies acutely to phylogenies (estimates of evolutionary relationships among species). Increasingly, phylogenies are estimated from increasingly large, genome-scale datasets using increasingly complex statistical methods  that require increasing levels of expertise and computational investment. 
Moreover, the resulting phylogenetic data provide an explicit historical perspective that critically informs research in a vast and growing number of scientific disciplines. 
One such use is the study of changes in rates of lineage diversification (speciation -- extinction) through time. 
As part of a meta-analysis in this area, we sought to collect phylogenetic data (comprising nucleotide sequence alignment and tree files) from $217$ studies published in $46$ journals over a $13$-year period. 
We document our attempts to procure those data (from online archives and by direct request to corresponding authors), and report results of analyses (using Bayesian logistic regression) to assess the impact of various factors on the success of our efforts. 
Overall, complete phylogenetic data for $\sim 60\%$ of these studies are effectively lost to science. 
Our study indicates that phylogenetic data are more likely to be deposited in online archives and/or shared upon request when: (1) the publishing journal has a strong data-sharing policy; (2) the publishing journal has a higher impact factor, and; (3) the data are requested from faculty rather than students. 
Although the situation appears dire, our analyses suggest that it is far from hopeless: recent initiatives by the scientific community---including policy changes by journals and funding agencies---are improving the state of affairs.

\newpage
\section*{Introduction}

Archiving and sharing published data is a social contract that is integral to the scientific enterprise \citep[][]{Vision_10}.
Sharing published data advances the scientific process by:
(1) exposing published results to independent verification (to identify errors and discourage fraud);
(2) providing the pedagogical material for educating students and training future researchers;
(3) acting as a test bed to guide the development of new methods, and; 
(4) providing a basis to identify and pursue new questions via synthesis/meta-analysis \citep[][]{Whitlock_11}.
Additionally, archiving published data protects our scientific investment, avoiding 
needless costs of data regeneration in terms of time, money, and environmental impact \citep[][]{Piwowar_11b}.

These considerations are particularly germane to phylogenetic data, which include both alignments (estimates of the positional homology of molecular sequences) and phylogenetic trees (estimates of the evolutionary relationships among species).
Phylogenetic trees for individual groups are inherently synthetic---combination of these `twigs' provides a natural approach for elucidating the entire Tree of Life \citep[{\it c.f.},][]{Maddison_07, opentree_14}.
Additionally, phylogenetic data have tremendous potential for reuse, often in ways that were completely unanticipated by the original studies: because they provide an explicit evolutionary perspective, phylogenies have become central to virtually all areas of research in evolutionary biology, ecology, molecular biology and epidemiology \citep[][]{Donoghue_2000, Piwowar_11, Stoltzfus_2012}.
Moreover, the generation of phylogenetic data is an increasingly arduous and technical enterprise.
Clearly, phylogenetic data are a precious scientific resource that must be preserved and shared in order to realize their full potential.   

The vast majority of phylogenies are estimated from molecular (primarily nucleotide) sequence data.
Although GenBank and similar public archives provide a robust \citep[albeit imperfect;][]{Noor_06} backstop against the complete loss of the \emph{raw} sequence data, these databases do not safeguard the associated \emph{phylogenetic} data: the alignments estimated from raw sequence data, and the trees inferred from those alignments.
Multiple sequence alignment---the process of estimating the positional homology of each nucleotide site comprising DNA sequences---is a difficult inference problem for which many approaches have been proposed \citep[][]{Notredame_2007, Thompson_2011}. 
Different algorithms (or different settings for a given algorithm) may yield dramatically different estimates of the alignment that, in turn, can substantially impact estimates of phylogeny \citep[][]{Wong_2008, Blackburne_2013}.
Moreover, the majority of phylogenetic studies are based on alignments that are subjected to `manual adjustment'  after being estimated using formal methods \citep[][]{Morrison_2009}, which effectively destroys the possibility of replicating published alignments from the corresponding raw sequence data.    
Even if the alignment could be dependably reproduced, replicating the published phylogeny requires a precise description of how the phylogenetic analysis was performed, details that are typically not provided in phylogenetic studies \citep[][]{Leebens-Mack_06}.  
Finally, even if the alignment and details of the analysis were available, re-generating the phylogeny remains a non-trivial proposition: the analysis of a single dataset may require hundreds or thousands of compute hours \citep[][]{Suchard_2009}.  

These issues have been appreciated for some time \citep[][]{Sanderson_93}, and motivated the development of a specialized online archive for phylogenetic data, TreeBASE \citep[][]{Sanderson_94}, more than 20 years ago.
Despite such noble efforts, it is increasingly evident that the loss of phylogenetic data is catastrophic:  
recent surveys estimate that $\sim 70\%$ of published phylogenetic data are lost forever \citep[][]{Drew_2013a, Drew_2013, Stoltzfus_2012}. 
In response to this crisis, several recent community initiatives have been proposed to encourage the preservation and sharing of phylogenetic data.
These include policy initiatives both by funding agencies (the NSF Data Management Plan established in 2011 that requires the preservation of data generated by funded research), and by journals/publishers \citep[the establishment of the Joint Data Archiving Policy, JDAP, by a consortium of prominent journals requiring the submission of data to online archives as a condition of publication;][]{Moore_2010, Whitlock_2010, Rausher_2010, Rieseberg_2010, Uyenoyama_2010},
and the establishment of a new online archive for evolutionary and ecological data, Dryad \citep[][]{Dryad_11}. 

We set out to perform a meta-analysis exploring the empirical prevalence of temporal changes in rates of lineage diversification.
To this end, we sought to collect the phylogenetic data from studies using the two most common statistical phylogenetic approaches for detecting temporal shifts in diversification rate; {\it i.e.}, the `gamma' statistic \citep[][`method 1']{Pybus_00} and the `birth-death likelihood' \citep[][`method 2']{rabosky_06} methods.  
To be included in our meta-analysis, we required two key data files from each published empirical study: (1) an alignment of nucleotide sequence data, and (2) an ultrametric tree (where the branch lengths are rendered proportional to relative or absolute time).
We document our attempts to procure these data (both via searches of online archives and by direct solicitation from the corresponding authors), and describe results of analyses exploring various factors associated with the availability of phylogenetic data.
We assess a number of correlates---the age of the study, the impact factor and data-sharing policy of the publishing journal, the status of the solicitor, etc.---with a focus on revealing the efficacy of recent community initiatives to ensure the preservation and promote the sharing of published phylogenetic data.

\section*{Methods}
In this section, we document our attempts to procure phylogenetic data from a large and random sample of studies exploring temporal variation in rates of lineage diversification published over a $13$-year period.
We first describe how we sought to collect these data, and then describe the analyses we performed to gauge the success of our efforts.

\subsection*{Data Collection}
During the months of October and November, 2013, we searched for articles citing the two methods papers
using the the ISI Web of Science cited-reference tool.
Our search identified a total of $470$ citing articles ($322$ and $148$ for methods $1$ and $2$, respectively).
Of these, a total of $217$ involved empirical analyses ($165$ and $52$ using methods $1$ and $2$, respectively).

For each study, we captured bibliometric data on authorship, publication month and year, and the name and impact factor of the publishing journal.
We also recorded the data-sharing policy of the publishing journal and whether it was a member of the JDAP initiative at the time of publication.
Specifically, we ascertained the data-sharing policy for each of the $46$ journals from the corresponding `instructions to authors' documentation (see Supporting Information).
Following \citep[][]{Piwowar_10}, we categorized journals that made \emph{no mention} of data sharing as having \emph{no policy}; those that \emph{encouraged} authors to share data upon publication were scored as having a \emph{weak policy}; those that \emph{required} data sharing as a condition of publication were scored as having a \emph{strong policy}; and those that were members of the JDAP initiative were scored as having \emph{JDAP membership}.
Finally, we noted whether the studies acknowledged funding support from the National Science Foundation (NSF). 

\begin{figure}[H]
\begin{center}
\includegraphics[width=3.5in]{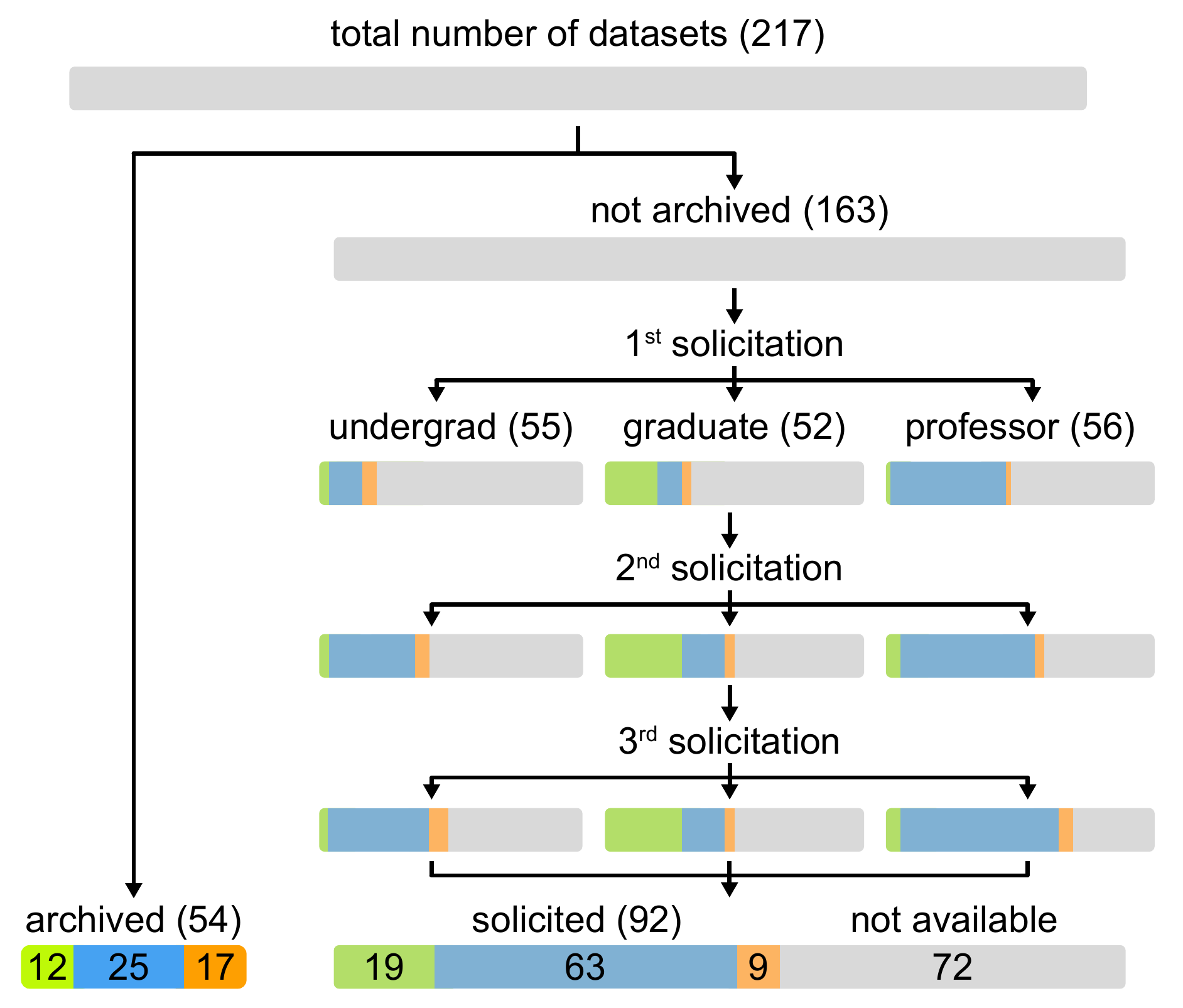}
\end{center}
\caption{
{\bf Flowchart of data acquisition.} We identified a total of $217$ articles exploring temporal variation in rates of lineage diversification. Data for $54$ of these studies were archived in online repositories; data for the remaining $163$ studies were solicited by direct requests to the corresponding author by an undergraduate student ($55$ studies), a graduate student ($52$), or a professor ($56$).
A maximum of three requests were made at weekly intervals.
Recovered phylogenetic data comprised tree files (green), alignment files (orange), or both (blue).
Datasets not obtained after the third request were deemed unavailable (gray).}
\label{flow_chart}
\end{figure}

For each study, we assessed whether data were available online by first searching each article for various keywords (``Dryad'', ``TreeBASE'', etc.), and pursued any links or references to archived data. 
If data could not be sourced directly from the article itself, we proceeded to examine any associated Supplemental Material files using a similar strategy.
Articles that did not submit their data to online repositories were targeted for direct solicitation using a semi-automated, multi-step approach (Figure \ref{flow_chart}).
Specifically, we wrote `templates' for three sequential messages comprising an initial, a followup, and a final request for published phylogenetic data.
In the messages, we identified ourselves, provided details of the requested data, and explained the reason for our request; 
that is, we explained that we were gathering data for a meta-analysis evaluating the prevalence of temporal changes in diversification rate, and we sought the sequence alignment and ultrametric tree files that were the used to assess temporal changes in diversification rates in the published study.

Each of the three message templates contained `fields' for several variables, including: the name and status of the solicitor; the name and email address of the corresponding author; and the year and title of the published article.
We divided the solicitations  evenly (and randomly) between the three of us.
This was intended both to share the burden equably, and also to assess any effect of the solicitor status, which comprised a professor (BRM), a graduate student (MRM) and an undergraduate student (AFM).
We then generated messages using \verb!R! scripts that populated the fields of the templates with the relevant information from the spreadsheet
(the templates and \verb!R! scripts are provided as Supporting Information).
Messages were sent at weekly intervals.
If we received a response, the corresponding author was precluded from receiving subsequent generic email messages, and we corresponded with them on an individual basis.
We recorded various details of each response, including whether the recipient sent the requested alignment file and/or tree file.  
Datasets not obtained at the end of this process were deemed unavailable.

A table summarizing information gathered for the $217$ studies is included as Supporting Information.
Following \citep[][]{Wicherts_11}, the data table has been anonymized to protect the identity of corresponding authors ({\it i.e.}, with regard to who did or did not archive and/or share phylogenetic data from published studies).
However, a key is available upon request to allow details of our analyses to be independently verified.
In any case, the issues that we document are general and should not be use to impugn the academic integrity of the individual researchers.

\subsection*{Data Analysis}
We used Bayesian logistic regression to explore correlations between data availability and several variables.
Under this approach, a  \emph{trial} is an attempt to recover data for a particular study either from online archives or by direct solicitation, which we deem a \emph{success} if we received data for that study.
The outcomes of a set of $n$ trials are contained in a data vector $\boldsymbol{x} = \{x_1,x_2,\ldots,x_n\}$, where $x_i$ is 1 if we obtained the relevant data for study $i$ and is 0 otherwise.
The outcome of each trial depends on a set of $k$ \emph{predictor variables} that may be continuous (\emph{e.g.}, the journal impact factor) or discrete (\emph{e.g.}, the status of the solicitor).
An $n \times k$ matrix $\mathcal{I}$, the \emph{design matrix}, describes the relationships between trials and predictor variables: $\mathcal{I}_{ij}$ is the value for predictor variable $j$ for trial $i$.
\emph{Parameters} relate the values of each predictor variable to the probability of success of each trial, and are described by the parameter vector $\boldsymbol\beta = \{\beta_1,\beta_2,\ldots,\beta_k\}$, where $\beta_i$ is the contribution of parameter $i$ to the probability of success.

In a Bayesian framework, we are interested in estimating the joint posterior probability distribution of the model parameters $\boldsymbol\beta$ conditional on the data $\boldsymbol{x}$. 
According to Bayes' theorem,
\begin{align*}
P(\boldsymbol\beta \mid \boldsymbol x) = \frac{ P( \boldsymbol{x} \mid \boldsymbol\beta  ) P(\boldsymbol\beta) }{ \int P( \boldsymbol{x} \mid \boldsymbol\beta  ) P(\boldsymbol\beta)\ \mathrm{d}\boldsymbol\beta },
\end{align*}
the \emph{posterior probability} of the model parameters, $P(\boldsymbol\beta \mid \boldsymbol x)$, is equal to \emph{likelihood} of the data given the model parameters, $P(\boldsymbol{x} \mid \boldsymbol\beta)$, multiplied by the \emph{prior probability} of the parameters, $P(\boldsymbol\beta)$, divided by the \emph{marginal likelihood} of the data.

Given the design matrix $\mathcal{I}$, the outcomes of each of the $n$ trials are conditionally independent, so that the likelihood of $\boldsymbol x$ is the product of the likelihoods for each individual trial:
\begin{align*}
P(\boldsymbol{x} \mid \boldsymbol\beta) = \prod_{i=1}^n P(x_i \mid \mathcal{I}, \boldsymbol\beta).
\end{align*}
The likelihood of observing the outcome of a particular trial is
\begin{align*}
P(x_i \mid \mathcal{I}, \boldsymbol\beta) = \begin{cases}
									\frac{1}{1+e^{-\omega_i} } & \text{if } x_i=1\\
									1-\frac{1}{1+e^{-\omega_i}} & \text{if } x_i=0,
								  \end{cases}
\end{align*}
where
\begin{align*}
\omega_i = \sum_{j=1}^{k} \mathcal{I}_{ij}\beta_j.
\end{align*}

We specified a multivariate normal prior probability distribution on the $\boldsymbol\beta$ parameters with means $\boldsymbol\mu$ and covariance matrix $\Sigma$.
The complexity of the marginal likelihood precludes an analytical solution to the posterior probability distribution.
Accordingly, we approximated the posterior probability distribution using the Markov chain Monte Carlo algorithm implemented in  the  \verb!R!  package \verb!BayesLogit! \citep[][]{Polson_13,R_13}.
This program uses conjugate prior and posterior probability distributions (via Polya-Gamma-distributed latent variables), which permits use of an efficient Gibbs sampling algorithm to approximate the joint posterior distribution of $\boldsymbol\beta$ conditional on the data.

We defined a set of predictor variables based on the bibliometric  metadata captured for each study. 
We included an \emph{intercept} predictor variable to describe the background probability of procuring data.
We treated \emph{age} ({\it i.e.}, months since publication) and \emph{journal impact factor} as continuous predictor variables, and \emph{journal policy}, \emph{NSF funding}, and \emph{solicitor status} as discrete predictor variables.
Discrete predictor variables for logistic regression are generally binary, assuming values of 0 or 1.
A few of our discrete bibliometric metadata, however, had more than two possible categories.
We therefore adopted an \emph{indicator-variable} approach in which predictor variables with $p$ categories are discretized into $p$ distinct indicators;  each study in a particular predictor category was then assigned a 1 for the corresponding indicator variable.
Under this approach, studies published in journals with no data-sharing policy were assigned a 1 for the \emph{no policy} variable, studies published in journals with a strong policy were assigned a 1 for the \emph{strong policy} variable, and studies published in journals that were members of the JDAP initiative at the time of publication were assigned a 1 for the \emph{JDAP membership} variable.
For the studies included in our direct-solicitation campaign, we also assigned values for solicitor status: datasets solicited by an undergraduate student were scored as 1 for the \emph{undergraduate student} variable, while those solicited by a professor were scored as 1 for the \emph{professor} variable.
In order to avoid overparameteriziation of the logistic model, we did not assign indicator variables for the \emph{weak-policy} or \emph{graduate-student} variables.
Accordingly, the values for \emph{no policy}, \emph{strong policy}, and \emph{JDAP membership} parameters are interpreted as effects relative to weak policies; similarly, the values for \emph{undergraduate student} and \emph{professor} parameters are interpreted as effects relative to a graduate student.
Details of the predictor variables and interpretations of the corresponding parameters are summarized in Table \ref{tab:paramtable}.

We analyzed various subsets of our data table in order to understand the relative importance of the predictor variables on different aspects of data availability.
Specifically, we defined subsets of our data table based on whether study data were sought: (1) by queries to online archives, (2) by direct solicitation from the corresponding author, or (3) either by queries to online archives \emph{or} by direct solicitation.
We further parsed our data table based on whether we successfully procured: (1) \emph{only} trees (\emph{i.e.}, the trial outcome was 1 if we acquired a tree and no alignment, and 0 otherwise); (2) \emph{only} alignments; (3) either alignments \emph{or} trees (\emph{i.e.}, the trial outcome was 0 if we acquired no data, and 1 otherwise), and; (4) both alignments \emph{and} trees (\emph{i.e.}, the trial outcome was 1 if we acquired both an alignment and a tree).
This defined 16 (overlapping) subsets of our data table.
Note that not all predictor variables apply to every subset of our data table; \emph{e.g.} the solicitor-status variable, \emph{undergraduate}, only applies to data that were directly solicited.
Details of the data subsets and their predictor variables are summarized in Table \ref{tab:modeltable}.

We estimated parameters for each data subset by performing four independent MCMC simulations, running each chain for $10^6$ cycles and saving every $100^{th}$ sample to reduce autocorrelation and file size. 
We assessed the performance of all MCMC simulations using the \verb!Tracer! \citep[][]{Drummond2012} and \verb!coda! \citep[][]{Plummer_2006} packages.
We monitored convergence of each chain to the stationary distribution by plotting the time series and calculating the Geweke diagnostic \citep[{\it GD}; ][]{Geweke_1992} for every parameter.
We assessed the mixing of each chain over the stationary distribution by calculating both the potential scale reduction factor \citep[{\it PSRF};][]{Gelman_1992} diagnostic and the effective sample size \citep[\emph{ESS};][]{Brooks_1997} for all parameters.
Values of all diagnostics for all parameters in all MCMC simulations indicate reliable approximation of the stationary (joint posterior probability) distributions: {\it e.g.}, $ESS >>1000$; $PSRF \approx 1$; {\it GD} $>>0.05$ (Tables \ref{tab:mcmc_archive_any}$-$\ref{tab:mcmc_combined_both}).
Additionally, we assessed convergence by comparing the four independent estimates of the marginal posterior probability density for each parameter, ensuring that all parameter estimates were effectively identical and SAE compliant \citep[{\it c.f.},][]{Brooks_1997}.
Based on these diagnostic analyses, we discarded the first $25\%$ of samples from each chain as burn-in, and based parameter estimates on the combined stationary samples from each of the four independent chains ($N = 30,000$).

\section*{Results and Discussion}
\begin{figure}[H]
\begin{center}
\includegraphics[width=6in]{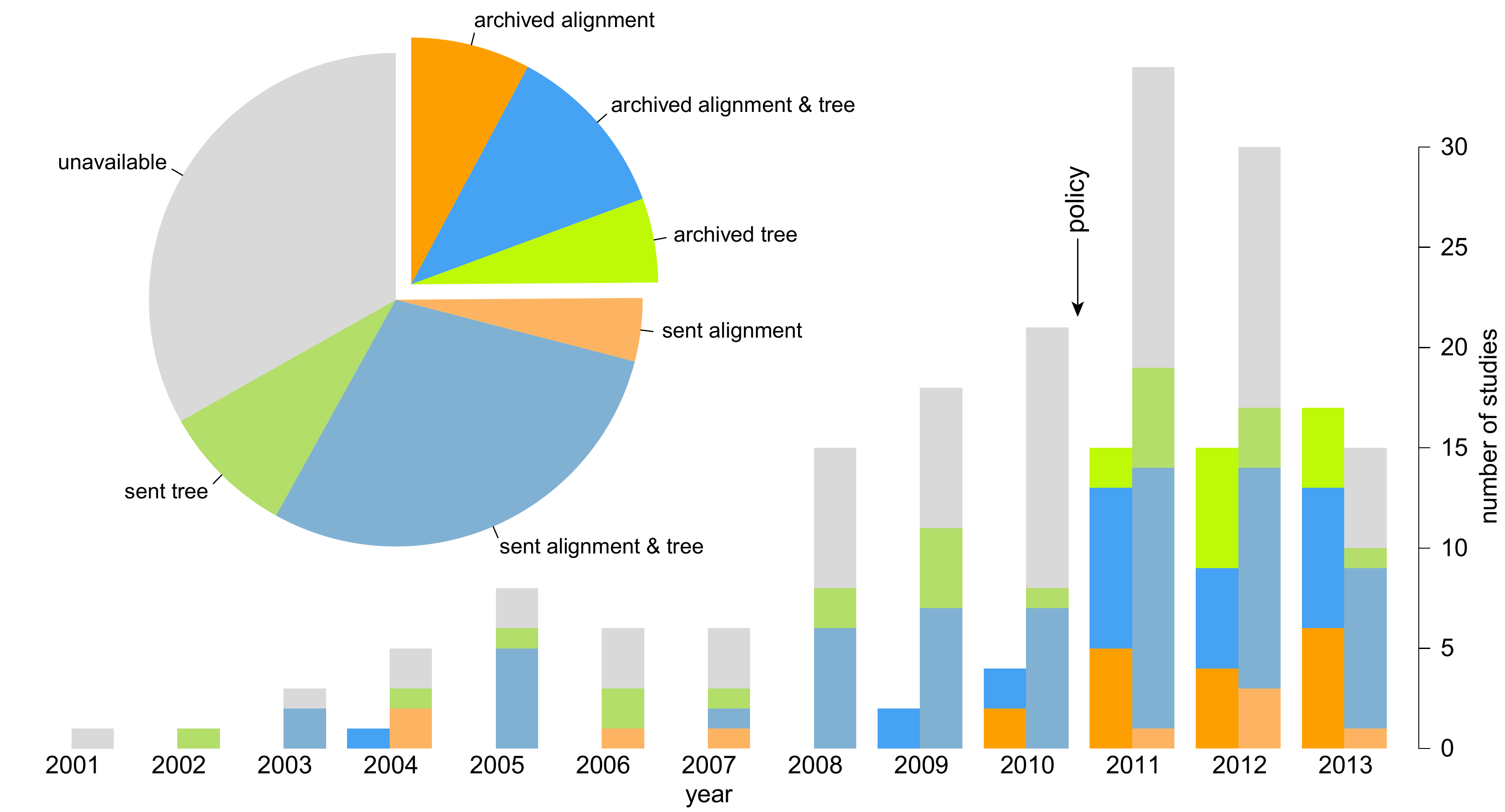}
\end{center}
\caption{
{\bf Detailed breakdown of data availability.} The number of studies with available phylogenetic data---as tree files (green), alignments files (orange) or both (blue), procured either from online archives or by direct request---organized by year of publication (barplot). Phylogenetic data of some kind (tree and/or alignment files) were available from an online archive for approximately $25\%$ of the studies, and additional data were successfully solicited by direct request for $42\%$ of the studies. Complete datasets were unavailable for $60\%$ of published studies, and data of any kind were unavailable for $33\%$ of studies (gray). The `policy' arrow indicates the onset of several community initiatives to improve the sharing and preservation of evolutionary (including phylogenetic) data, which coincides with a marked increase in the deposition of phylogenetic data to online archives.}
\label{pie_bar}
\end{figure}

Overall, our efforts secured complete phylogenetic data for $\sim 40\%$ of the published studies (Figure \ref{pie_bar}).
Accordingly, invaluable phylogenetic data for more than half of these studies are effectively lost to science. 
From online archives, we successfully procured \emph{complete} phylogenetic data (both the tree and alignment files) for $11.5\%$ of the studies, and \emph{partial} datasets (either the tree or alignment files) for an additional $13.4\%$ of the studies were archived: $5.5\%$
of these cases had only tree files, $7.9\%$
had only alignment files.
Of these online accessions, $24$ were archived in Dryad, $22$ in TreeBASE, and $8$ as supplemental files on journal websites.
Our (in)ability to recover phylogenetic datasets from online archives over the \emph{entire} 13-year period is comparable to that of recent reports regarding phylogenetic data---where archival rates range from $\sim 4\% - 16.7\%$ \citep[][]{Hughes_2011, Stoltzfus_2012, Drew_2013a}---and also falls within the scope of archival rates for non-phylogenetic data, which range from $\sim 14\%-48\%$ \citep[][]{Piwowar_07, Alsheikh-Ali_11, Vines_13}.
However, our results also reveal a dramatic increase in the archiving of phylogenetic data since 2011; {\it e.g.}, datasets from more than half of the studies published in 2013 were deposited in online archives (Figure \ref{pie_bar}). 

Our direct-solicitation campaign entailed the exchange of $786$ emails over the course of four weeks (BRM: $n = 341$; MRM: $n = 212$; AFM: $n =233$).
We received responses to $61.3\%$ of the $163$ messages we sent to corresponding authors ($37\%$, $18\%$, and $7\%$ after the first, second and third message, respectively), $38.7\%$ of the authors never responded to any messages ($28\%$, $46\%$, and $42\%$ for BRM, MRM, and AFM, respectively).
Although $20.2\%$ of the messages were initially undeliverable (owing to invalid/obsolete email addresses), we were able to resolve contact information for all but $3\%$ of the corresponding authors (by performing Internet searches and/or contacting study co-authors).
Our $61\%$ response rate is comparable to that of previous studies.
A recent survey \citep[][]{Drew_2013a} reported a $40\%$ response rate to direct requests for phylogenetic data, which falls within the range for studies involving non-phylogenetic data: {\it e.g.}, $20\%$ for medical/clinical trial data \citep[][]{Savage_09};  $27\%$ for psychological trial data \citep[][]{Wicherts_06}; and $71\%$ for population-genetic data \citep[][]{Vines_13}.   

By directly contacting corresponding authors, we successfully procured complete phylogenetic datasets for $29.0\%$
of the published studies, and partial datasets for an additional $12.9\%$
of the studies: $8.8\%$ 
of corresponding authors sent only tree files, and $4.1\%$ 
sent only alignment files (Figure \ref{pie_bar}).
Our success in procuring complete $(29\%)$ or some form $(42\%)$ of phylogenetic data by direct solicitation compares favorably to that of a recent study \citep[$16\%$;][]{Drew_2013a}, but again is within the range reported for non-phylogenetic data; {\it e.g.}, $10\%$ for medical/clinical trial data \citep[][]{Savage_09}; $26\%$ for psychological-trial data \citep[][]{Wicherts_06}; $45\%$ for gene-expression data \citep[][]{Piwowar_11a}; $48\%$ for cancer microarray data \citep[][]{Piwowar_07}; $59\%$ for population-genetic data \citep[][]{Vines_13}.

\begin{figure}[H]
\begin{center}
\includegraphics[width=6in]{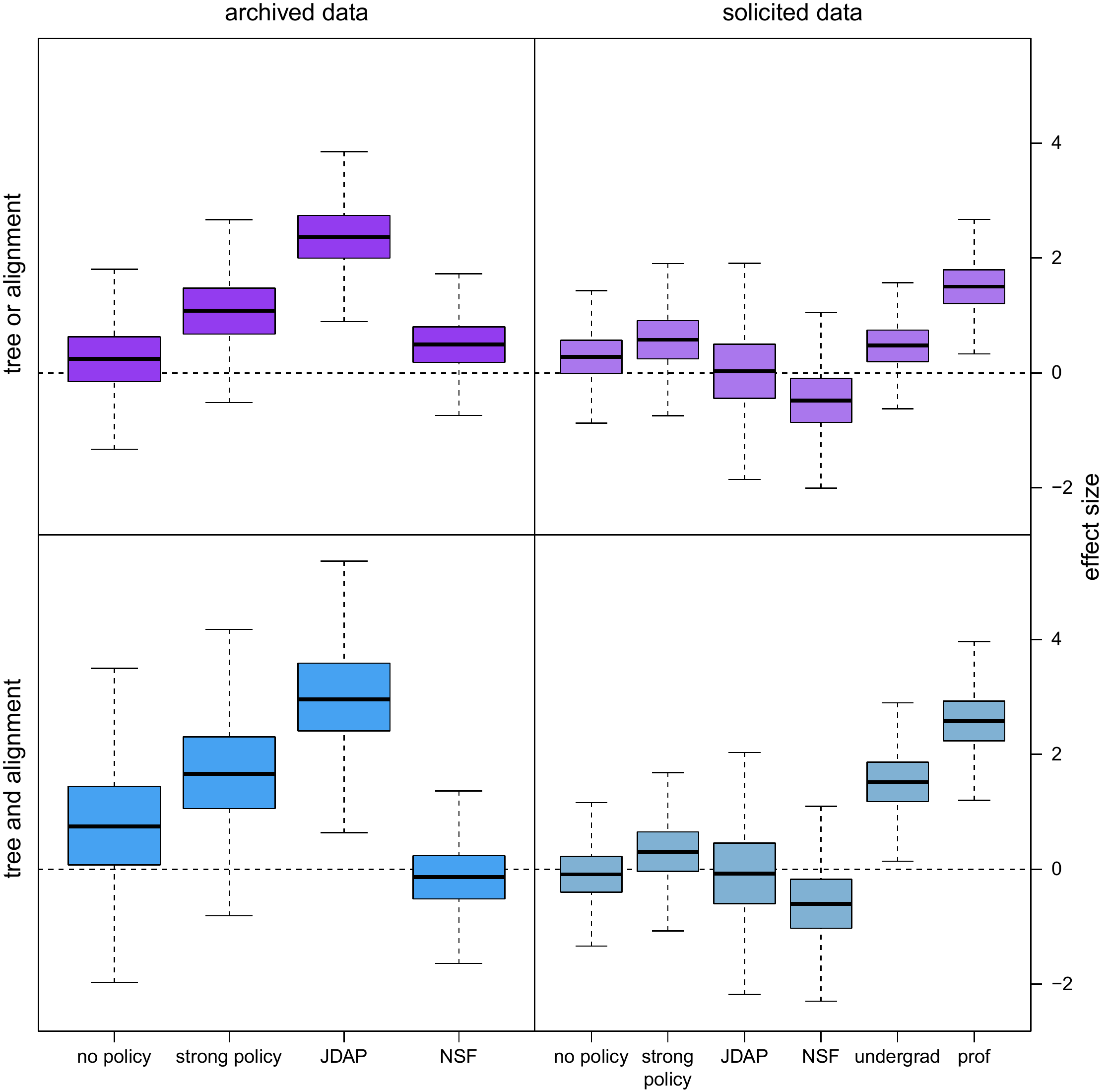}
\end{center}
\caption{
{\bf Correlates of data availability.} We used Bayesian logistic regression to estimate the effect of several variables on the on the probability that phylogenetic datasets were either available from a public archive (left column) or could be successfully procured by direct solicitation (right column).
Specifically, for all datasets we explored the effect of the data-sharing policy of the publishing journal (scored as \emph{none}, \emph{weak}, \emph{strong}, or \emph{JDAP membership}) and the impact of funding-agency policy (\emph{NSF}).
For solicited datasets, we also assessed the impact of solicitor status (\emph{undergraduate}, \emph{graduate}, or \emph{professor}).
We estimated effects of these variables on our ability to successfully procure \emph{either} the tree or alignment files (top panels), or \emph{both} the tree and alignment files (bottom panels) for a given study. 
The estimated effect size for a given variable reflects its contribution to the probability of successfully acquiring the data. 
For each variable, the marginal distribution of its estimated effect size is summarized as a boxplot, indicating the median effect (solid line), $\pm 1$ interquartile range (box), and $1.5$ interquartile range (whisker) of the corresponding posterior probability distribution. 
Journal-policy effects are relative to the effect of a weak policy, and solicitor-status effects are relative to that of graduate student.
The predictor variables and interpretation of the corresponding parameters are described in Table \ref{tab:paramtable}.}
\label{boxplots}
\end{figure}

The results of our logistic-regression analysis provide insights into factors associated with the availability of published phylogenetic data (Figure \ref{boxplots}; Tables \ref{tab:results_archive_any_relative_probs}$-$\ref{tab:results_sent_any_relative_probs}).
Studies published in journals with strong data-sharing policies are more likely to archive both complete (tree and alignment files) and incomplete (tree or alignment files) phylogenetic data, and are also more likely to provide complete and incomplete phylogenetic data upon direct request.
Strikingly, the availability of phylogenetic data (via online archives or direct solicitation) from studies published in journals with weak data-sharing policies is comparable to (or slightly worse) than that of studies published in journals with no data-sharing policy \citep[{\it c.f.}, ][]{Piwowar_10, Vines_13}.
This observation substantiates recent calls for establishing strong (and stringently enforced) data-sharing policies \citep[][]{Savage_09, Piwowar_10, Whitlock_11, Drew_2013, Drew_2013a}.
The efficacy of such policies is evident for studies published in JDAP journals.
Surprisingly, there is a \emph{low} probability of directly soliciting data for studies published in JDAP journals.
However, this likely reflects the fact that the data from these studies are so often available in online archives that there is essentially no \emph{need} for direct solicitation. 

Our analyses also indicate that corresponding authors are more likely to grant data requests from faculty than from students (Figure \ref{boxplots}).
This may simply reflect the fact that the faculty solicitor (BRM) is acquainted with a larger proportion of the corresponding authors. 
However, this does not explain why corresponding authors are more likely to provide data to undergraduate than to graduate students.
An alternative (but not mutually exclusive) explanation involves the perceived risks of data sharing.
Authors may be reluctant to share published data for fear (reasonable or not) that reanalysis may identify errors and/or reach contradictory conclusions
\citep[][]{Ceci_1983, Nature_2006}.
This idea has, in fact, been substantiated by a recent study demonstrating that reluctance to share published data is significantly correlated with weaker evidence and a higher prevalence of apparent errors in the reporting of statistical results \citep[][]{Wicherts_11}.
Accordingly, corresponding authors may perceive requests from undergraduate students to present less potential risk than those from graduate students, whereas the potential risks presented by faculty requests are balanced by their greater familiarity to the authors.

\begin{figure}[H]
\begin{center}
\includegraphics[width=6in]{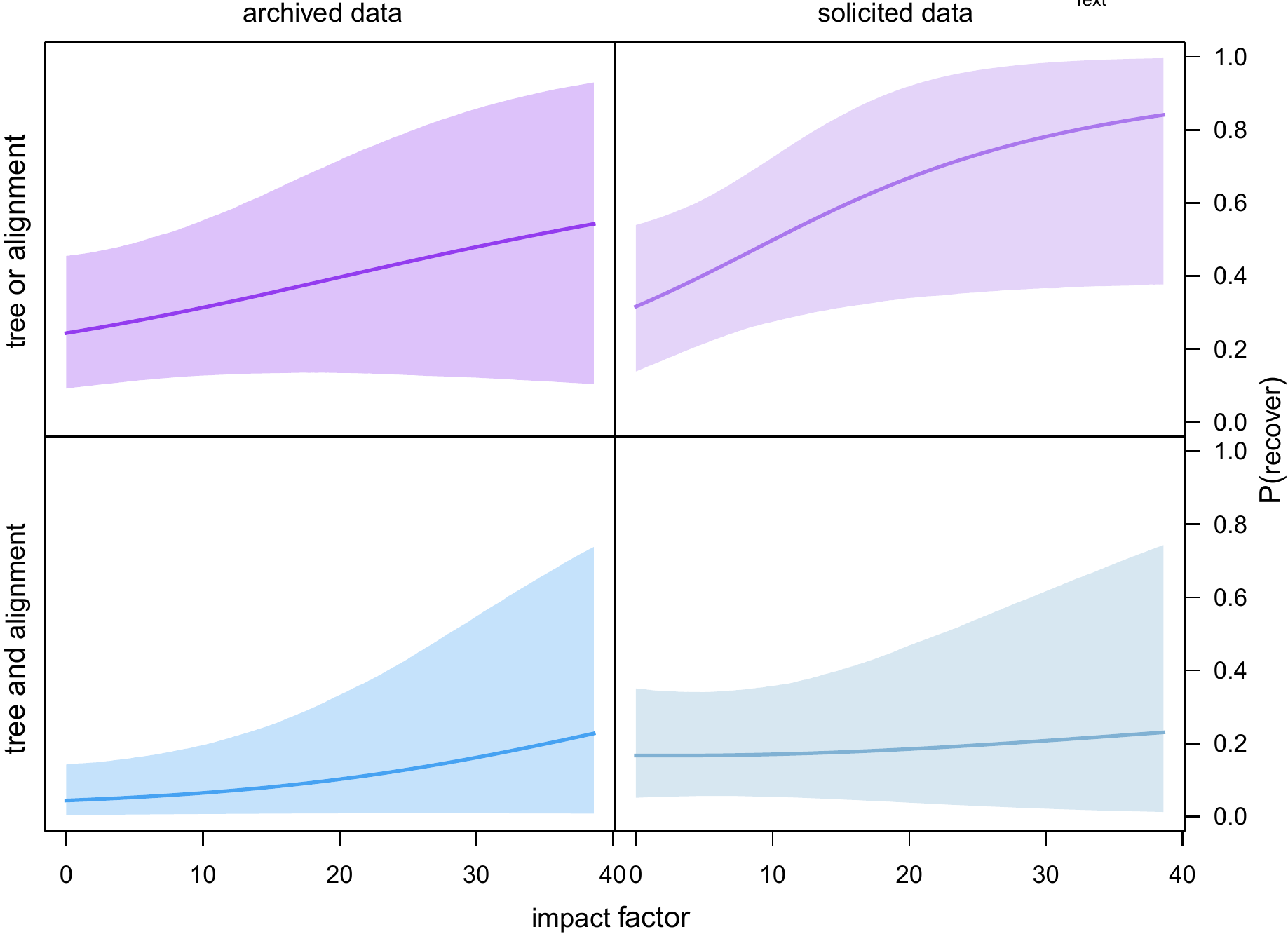}
\end{center}
\caption{
{\bf Availability of phylogenetic data as a function of impact factor.} We estimated the effect of the impact factor of the publishing journal on our ability to procure partial (top panels) and complete (bottom panels) phylogenetic datasets from online archives (left panels) or by direct solicitation (right panels).
Generally, studies published in journals with a higher impact factor are more likely to both deposit the corresponding (partial or complete) datasets in online archives and to provide those data upon direct request.
The shaded areas reflect the $95\%$ credible intervals of the estimates.}
\label{recovery_IF}
\end{figure}

The influence of journal impact factor on data availability might also be interpreted from the perspective of perceived risk.
As for non-phylogenetic data \citep[][]{Piwowar_10, Vines_13}, our analyses indicate that studies published in journals with a higher impact factor are more likely to both deposit their phylogenetic data in online archives and provide these data upon direct request (Figure \ref{recovery_IF}).
If willingness to share published data is correlated with the quality of the research \citep[][]{Wicherts_11}, and if research quality is correlated with the impact factor of the publishing journal, then journal impact factor should positively predict data availability.
An alternative (perhaps less conspiratorial) explanation for the correlation between journal impact factor and data availability invokes an indirect effect of journal impact factor on journal data-sharing policy.
That is, by virtue of their greater prestige, journals with higher impact factors may have greater reign to impose stronger (and more strictly enforced) data-sharing policies on contributing authors \citep[][]{Vines_13}.

\begin{figure}[H]
\begin{center}
\includegraphics[width=6in]{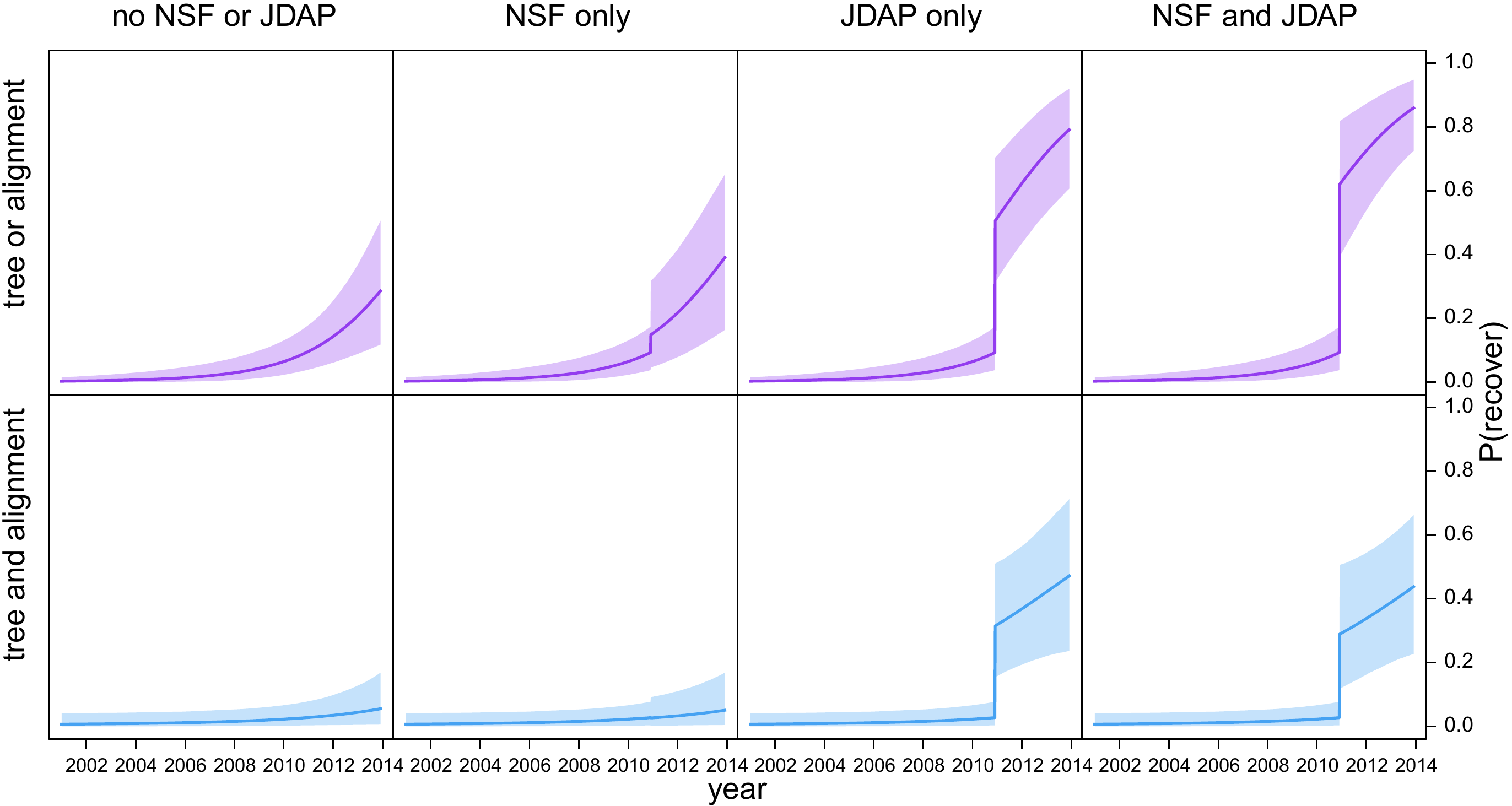}
\end{center}
\caption{
{\bf Availability of archived phylogenetic data as a function of age.} We estimated the effect of publication age on our ability to procure partial (top panels) and complete (bottom panels) phylogenetic datasets from online archives. 
Overall, the probability of recovering archived phylogenetic data increases toward the present, with a conspicuous recent increase for partial datasets (left panels).
The recent surge of archived phylogenetic data likely reflects recent policy changes (middle panels): 
studies with NSF funding are more likely to archive alignment (but not tree) files ({\it c.f.}, Table \ref{tab:results_archive_alignments}); whereas studies published in journals with JDAP membership are dramatically more likely to archive both partial and complete phylogenetic datasets.
The effects of these policy initiatives are not strictly additive (right panels): the correlation of these predictor variables suggests that studies published in JDAP journals are likely to have NSF funding.  
Shaded areas reflect the $95\%$ credible intervals.}
\label{recovery_age_archive}
\end{figure}

\begin{figure}[H]
\begin{center}
\includegraphics[width=6in]{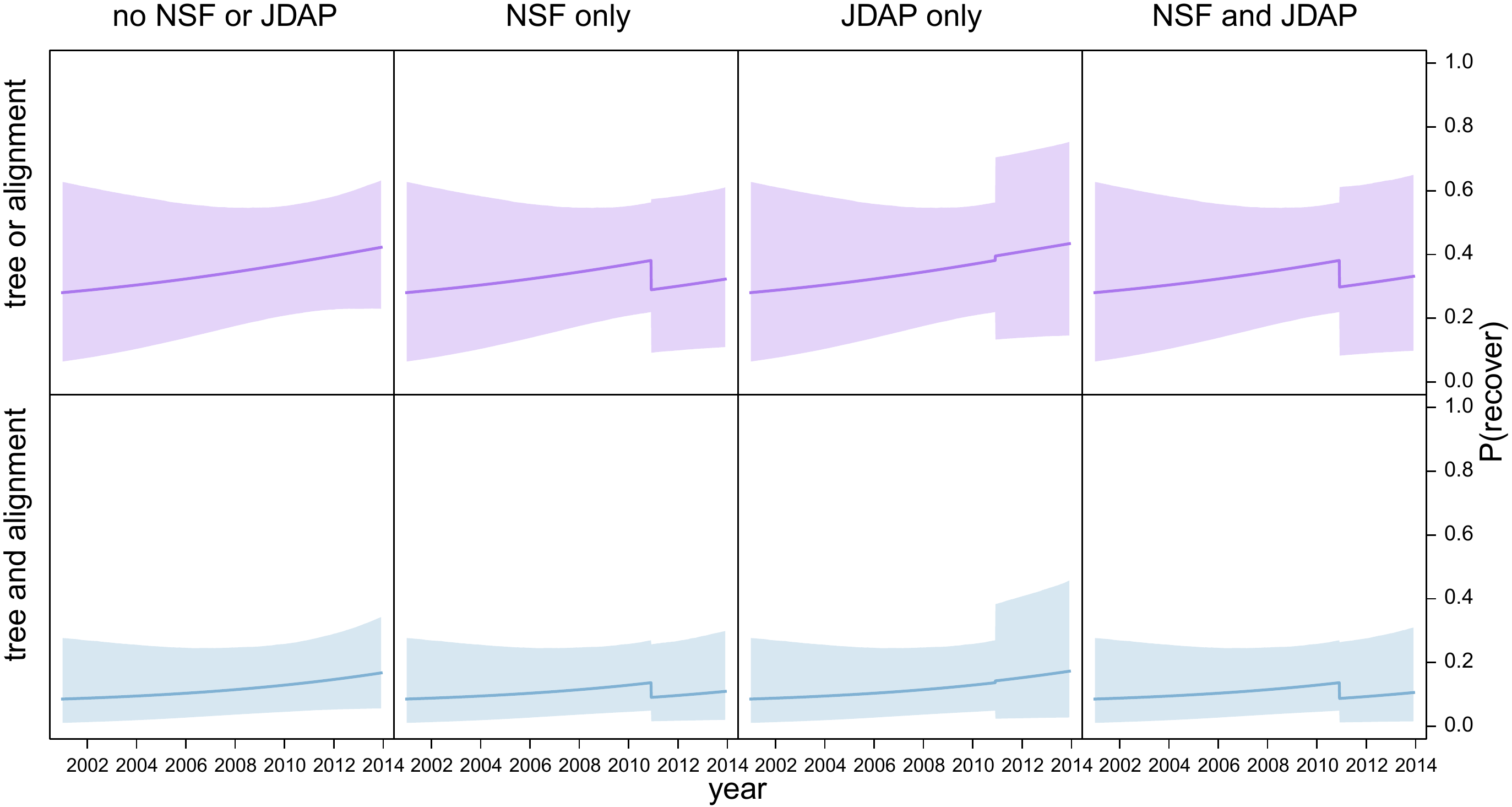}
\end{center}
\caption{
{\bf Availability of solicited phylogenetic data as a function of age.} We estimated the effect of publication age on our ability to procure partial (top panels) and complete (bottom panels) phylogenetic datasets by direct solicitation. 
Overall, the probability of successfully recovering phylogenetic data decreases over time (left panel).
Paradoxically, the probability of soliciting data from studies with NSF funding and/or published in JDAP journals is \emph{lower} than that for studies without NSF funding and/or published in non-JDAP journals. 
However, this likely reflects the fact that the data from these studies are so often available in online archives that there is essentially no \emph{need} for direct solicitation.
Shaded areas reflect the $95\%$ credible intervals.}
\label{recovery_age_solicit}
\end{figure}

As in previous studies \citep[][]{Evangelou_2005}, our results indicate that data availability decreases markedly over time.
Several corresponding authors reported that the requested datasets had been misplaced or had been lost due to hard-drive failures.
As noted above, there appears to be a distinct uptick in the availability of data from studies published since $2011$; this trend was particularly pronounced for archived data (Figure \ref{recovery_age_archive}).
This pattern may simply indicate that the decay of archived phylogenetic data is nonlinear.
Our findings, however, indicate that the recent surge in archived phylogenetic data is attributable to policy changes.
Studies with NSF funding are $\sim 1.4$ times more likely to archive some kind of phylogenetic data (tree or alignment files), but are actually \emph{less} likely to archive complete phylogenetic data (Table \ref{tab:results_archive_any_relative_probs}).
Curiously, the NSF mandate has led to a drastic increase in archiving alignment (but not tree) files  (Table \ref{tab:results_archive_alignments}).
By contrast, studies published in journals with JDAP membership are $\sim 2.8$ and $\sim 8.6$ times more likely to archive partial and complete phylogenetic datasets, respectively (Table \ref{tab:results_archive_any_relative_probs}; Figure \ref{recovery_age_archive}).
Paradoxically, the probability of successfully soliciting data from studies with NSF funding and/or published in JDAP journals is \emph{lower} than that for studies without NSF funding and/or published in non-JDAP journals (Figure \ref{recovery_age_solicit}). 
However, this likely reflects the decreased demand for these data by direct solicitation.

\section*{Summary}
Phylogenetic data are a precious scientific resource: molecular sequence alignments and phylogenies are expensive to generate, difficult to replicate, and have seemingly infinite potential for synthesis and reuse.
At face value, our results support the conclusion of recent studies \citep[][]{Drew_2013a, Drew_2013, Stoltzfus_2012} that the loss of phylogenetic data is catastrophic:
complete phylogenetic datasets have been lost for $\sim 60\%$ of the studies we surveyed.
Our results also identify factors associated with (phylogenetic) data availability that have been implicated by previous studies:
the probability of procuring phylogenetic data is strongly predicted the age of the study, and the data-sharing policy and impact factor of the publishing journal.

Unlike previous studies, however, our survey of phylogenetic datasets spans important policy initiatives and infrastructural changes, and so provides an opportunity to assess the efficacy of those recent measures.
Overall, the positive impact of these community initiatives has been both substantial and immediate.
Even at this very early stage---spanning the first three years since the introduction of these policies---the archival rate of phylogenetic data has increased dramatically.
Specifically, the proportion of studies that archived partial or complete phylogenetic data since 2011 has increased $4.8$-fold and $2.9$-fold, respectively.
Moreover the proportion of archived phylogenetic data has increased each year since the policy changes, and 
deposition rates of phylogenetic data to Dryad have been $4.3$ times that of the more established TreeBASE archive.
The prospects for future progress along these lines appear promising: membership of the JDAP consortium has almost tripled in the three years since its formation.

Although recent policy initiatives have had a clear and welcome effect on the preservation and sharing of phylogenetic data, there nevertheless remains considerable scope for improvement.
The NSF data-management policy, for example, has increased the preservation of alignments but not phylogenetic trees. 
This is unfortunate, both because phylogenies are more computationally expensive than alignments, and also because most of the reuse of phylogenetic data entails trees rather than sequence alignments \citep[][]{Piwowar_11, Stoltzfus_2012}.
Moreover, although relative archival rates have increased dramatically, the absolute rate remains low: despite recent policy initiatives, a large proportion of datasets are not being captured in online archives. 
Sustaining the momentum of recent initiatives could be achieved via small measures that increase the benefits and decrease the costs of data sharing to data generators. 
Although authors who archive data are rewarded with increased citation rates \citep[][]{Piwowar_07, Piwowar_13}, this incentive could be enhanced by rewarding the collection of data as an achievement in its own right.
Journal policies can encourage the direct citation of archived datasets in addition to the studies in which the data were generated, 
and funding agencies and academic institutions can recognize alternative metrics that acknowledge the scientific value of data \citep[][]{Piwowar_13a}. 
Concordantly, the perceived costs of data sharing could be reduced by implementing more flexible embargo policies that protect the priority access of data generators
\citep[][]{Vision_10, Roche_2014}.

Clearly, we have a long way to go in order to adequately preserve and freely share phylogenetic data, and the road ahead will not be easy. 
Nevertheless, our findings suggest that we are moving in the right direction; we are beginning to glimpse the dawn of open access to phylogenetic data.

\section*{Acknowledgments}
We are grateful to Bob Thomson for sharing insights on this work, and to all of the corresponding authors for sharing the requested phylogenetic datasets.
This research was supported by NSF grants DEB-0842181 and DEB-0919529 awarded to BRM.

\newpage
\bibliography{tree_hunting}

\newpage

\newpage

\newpage

\newpage

\newpage

\newpage

\newpage

\newpage
\newpage
\section*{Tables}

\begin{table}[H]
\caption{\bf{Summary of logistic model parameters and their interpretation}} \label{tab:paramtable}
\begin{tabular}{llp{9cm}}
\toprule
Parameter & Predictor variable & Interpretation \\
\midrule
$\beta_\text{I}$ & \emph{intercept} & The ``base'' logs-odds of retrieving the data, irrespective. of other model parameters. \\
\rowcolor{gray!25}
$\beta_\text{age}$ & \emph{age} & The change in log-odds of retrieving the data per month of the study's age. \\
$\beta_\text{IF}$ & \emph{impact factor} & The change in log-odds of retrieving the data per unit impact factor of the journal in which the study was published. \\
\rowcolor{gray!25}
$\beta_\text{none}$ & \emph{no policy} & The change in log-odds of retrieving the data if the study was published in a journal with no data-availability policy (relative to a weak policy). \\
$\beta_\text{strong}$ & \emph{strong policy} & The change in l	og-odds of retrieving the data if the study was published in a journal with a strong data-availability policy (relative to a weak policy). \\
\rowcolor{gray!25}
$\beta_\text{JDAP}$ & \emph{JDAP membership} & The change in log-odds of retrieving the data if the study was published in a member of the JDAP initiative beginning 2011 (relative to a weak policy). \\
$\beta_\text{NSF}$ & \emph{NSF funding} & The change in log-odds of retrieving the data if the study reported NSF funding beginning 2011. \\
\rowcolor{gray!25}
$\beta_\text{undergrad}$ & \emph{undergraduate student} & The change in log-odds of retrieving the data if it was solicited by an undergraduate student (relative to a graduate student). \\
$\beta_\text{prof}$ & \emph{professor} & The change in log-odds of retrieving the data if it was solicited by a professor (relative to a graduate student). \\
\rowcolor{gray!25}
$\beta_\text{solicited}$ & \emph{solicited} & The change in log-odds of retrieving the data if it was solicited (relative to archived). \\
\end{tabular}
\end{table}

\begin{table}[H]
\caption{\bf{Relative probability of obtaining phylogenetic data from online archives.}}
\centering
\begin{tabular}{lcccccc}
\toprule
& \multicolumn{3}{c}{alignments or trees} & \multicolumn{3}{c}{alignments and trees} \\
\cmidrule(r){2-4} \cmidrule(r){5-7}
& & \multicolumn{2}{c}{$95\%$ HPD} & & \multicolumn{2}{c}{$95\%$ HPD} \\
\cmidrule(r){3-4} \cmidrule(r){6-7}
& mean & lower & upper & mean & lower & upper \\
\midrule
\emph{no policy} & 1.16982 & 0.41697 & 2.34887 & 1.86843 & 0.05448 & 10.95330 \\ 
\rowcolor{gray!25}
\emph{strong policy} & 1.83196 & 0.78871 & 3.52417 & 3.91670 & 0.32121 & 21.84066 \\ 
\emph{JDAP membership} & 2.76346 & 1.40420 & 5.45642 & 8.58101 & 1.86761 & 54.18778  \\ 
\rowcolor{gray!25}
\emph{NSF funding} & 1.37201 & 0.66552 & 2.32973 & 0.91031 & 0.20170 & 2.09082 \\ 
\label{tab:results_archive_any_relative_probs}
\end{tabular}
\end{table}

\begin{table}[H]
\caption{\bf{Relative probability of procuring phylogenetic data by solicitation.}}
\centering
\begin{tabular}{lcccccc}
\toprule
& \multicolumn{3}{c}{alignments or trees} & \multicolumn{3}{c}{alignments and trees} \\
\cmidrule(r){2-4} \cmidrule(r){5-7}
& & \multicolumn{2}{c}{$95\%$ HPD} & & \multicolumn{2}{c}{$95\%$ HPD} \\
\cmidrule(r){3-4} \cmidrule(r){6-7}
& mean & lower & upper & mean & lower & upper \\
\midrule
\emph{no policy} & 1.15764 & 0.65935 & 1.78039 & 0.93791 & 0.34727 & 1.81923 \\ 
\rowcolor{gray!25}
\emph{strong policy} & 1.31740 & 0.77646 & 2.08017 & 1.31175 & 0.43575 & 2.49202 \\ 
\emph{JDAP membership} & 1.02670 & 0.29913 & 1.84746 & 1.02594 & 0.09284 & 2.52197 \\ 
\rowcolor{gray!25}
\emph{NSF funding} & 0.76364 & 0.26648 & 1.34441 & 0.65328 & 0.11797 & 1.40329 \\ 
\emph{undergraduate student} & 1.26807 & 0.79462 & 1.92777 & 2.76418 & 1.15683 & 6.10738 \\ 
\rowcolor{gray!25}
\emph{professor} & 1.78462 & 1.18732 & 2.82218 & 4.21124 & 1.79958 & 9.57003 \\ 
\label{tab:results_sent_any_relative_probs}
\end{tabular}
\end{table}

\newpage
\setcounter{section}{0}
\setcounter{table}{0}
\setcounter{figure}{0}
\setcounter{equation}{0}
\renewcommand{\thesection}{SI.\arabic{section}}
\renewcommand{\theequation}{S.\arabic{equation}}

\section*{Supporting Information}
\renewcommand{\thetable}{S.\arabic{table}}

\subsection*{Example template messages used for direct solicitation of phylogenetic datasets}

\subsubsection*{Undergraduate First Request}

Dear Dr. $<$\verb!corresponding author!$>$,

\medskip
\noindent
My name is $<$\verb!solicitor!$>$, and I'm an undergraduate researcher working in the Moore lab at UC Davis. We are doing a meta-analysis of density-dependent rates of lineage diversification. We would like to include your $<$\verb!publication year!$>$ study $<$\verb!study title!$>$ in our meta-analysis, and were hoping that you could send us any tree files that you used to perform the diversification-rate analysis in your study, as well as the alignment used to generate the study tree.

\noindent
Thank you very much for your time and consideration.

\medskip
\noindent
Sincerely,
$<$\verb!solicitor!$>$

\subsubsection*{Undergraduate Second Request}

Dear Dr. $<$\verb!corresponding author!$>$,

\medskip
\noindent
I emailed you a few days ago asking for your tree files and alignments from your $<$\verb!publication year!$>$ study, $<$\verb!study title!$>$, for inclusion in a meta-analysis of density-dependent lineage diversification. Since I have not yet heard back from you, I would like to reiterate my request for the data, which we would very much like to include.

\medskip
\noindent
Sincerely,
$<$\verb!solicitor!$>$

\subsubsection*{Undergraduate Third Request}

Dear Dr. $<$\verb!corresponding author!$>$,

\medskip
\noindent
I emailed you some time ago asking for your tree file and alignment from your $<$\verb!publication year!$>$ study, $<$\verb!study title!$>$, for inclusion in a meta-analysis of density-dependent lineage diversification. I would still appreciate it if you could send me the relevant data.

\medskip
\noindent
Sincerely,
$<$\verb!solicitor!$>$

\newpage
\subsection*{Example R script used to generate messages for direct-solicitation campaign}
\begin{figure}[!ht]
\begin{center}
\includegraphics[width=6in]{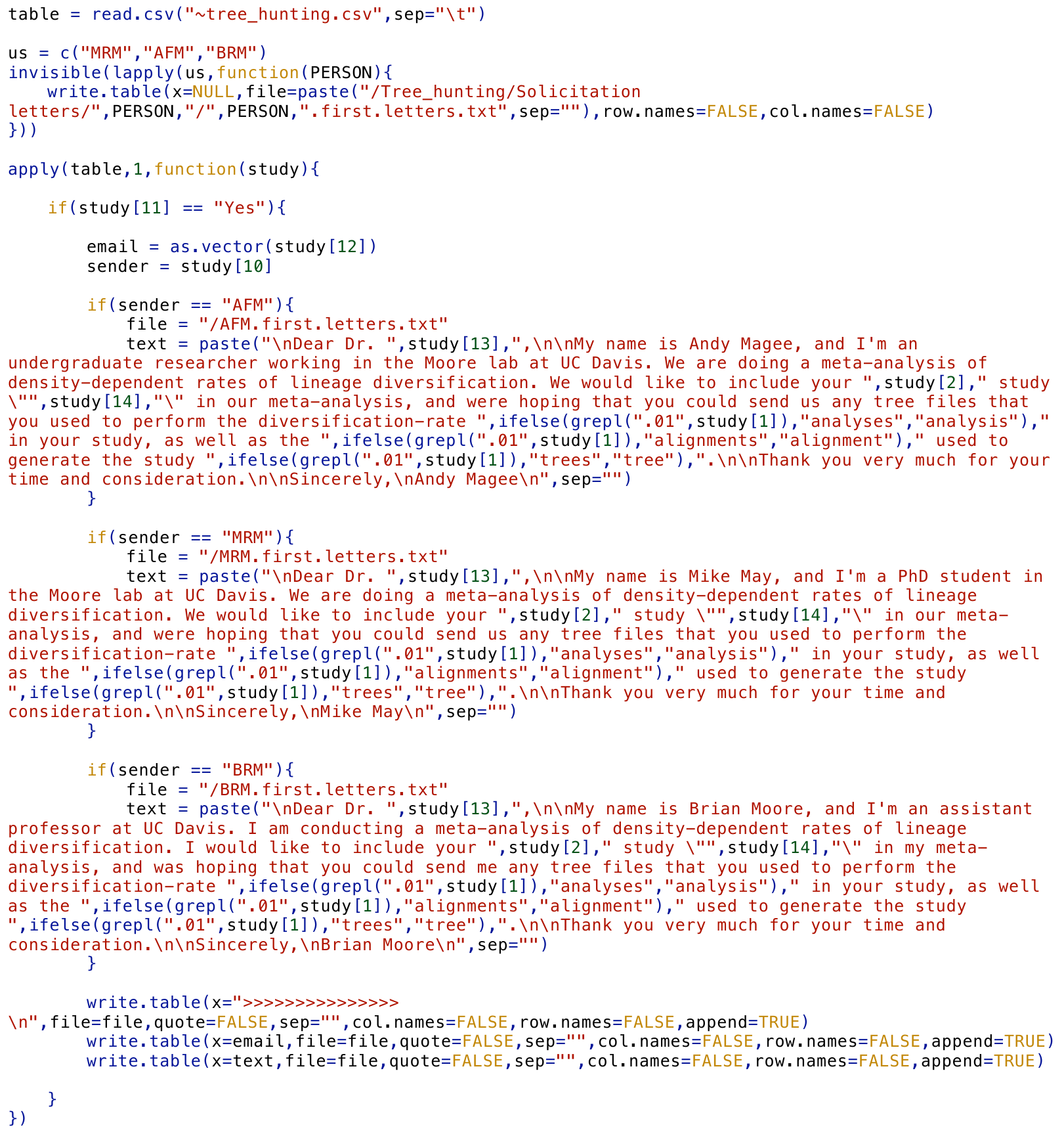}
\end{center}
\end{figure}

\begin{table}
\caption{\bf{Summary of datasets and models.}} \label{tab:modeltable}
\begin{tabular}{lllp{8.8cm}}
\toprule
Name & Source & Data & Predictor variables \\
\midrule
$x_\text{a,a}$ & archived & alignments only & \emph{intercept, age, impact factor, no policy, strong policy, JDAP membership, NSF funding} \\ 
\rowcolor{gray!25}
$x_\text{a,t}$ & archived & trees only & \emph{intercept, age, impact factor, no policy, strong policy, JDAP membership, NSF funding} \\ 
$x_\text{a,e}$ & archived & alignments or trees & \emph{intercept, age, impact factor, no policy, strong policy, JDAP membership, NSF funding} \\ 
\rowcolor{gray!25}
$x_\text{a,b}$ & archived & alignments and trees & \emph{intercept, age, impact factor, no policy, strong policy, JDAP membership, NSF funding} \\ 
$x_\text{s,a}$ & solicited & alignments only & \emph{intercept, age, impact factor, no policy, strong policy, JDAP membership$^*$, NSF funding$^*$, undergraduate student, professor} \\ 
\rowcolor{gray!25}
$x_\text{s,t}$ & solicited & trees only & \emph{intercept, age, impact factor, no policy, strong policy, JDAP membership, NSF funding, undergraduate student, professor} \\ 
$x_\text{s,e}$ & solicited & alignments or trees & \emph{intercept, age, impact factor, no policy, strong policy, JDAP membership, NSF funding, undergraduate student, professor} \\ 
\rowcolor{gray!25}
$x_\text{s,b}$ & solicited & alignments and trees & \emph{intercept, age, impact factor, no policy, strong policy, JDAP membership, NSF funding, undergraduate student, professor} \\ 
$x_\text{c,a}$ & combined & alignments only & \emph{intercept, age, impact factor, no policy, strong policy, JDAP membership, NSF funding, solicited} \\ 
\rowcolor{gray!25}
$x_\text{c,t}$ & combined & trees only & \emph{intercept, age, impact factor, no policy, strong policy, JDAP membership, NSF funding, solicited} \\ 
$x_\text{c,e}$ & combined & alignments or trees & \emph{intercept, age, impact factor, no policy, strong policy, JDAP membership, NSF funding, solicited} \\ 
\rowcolor{gray!25}
$x_\text{c,b}$ & combined & alignments and trees & \emph{intercept, age, impact factor, no policy, strong policy, JDAP membership, NSF funding, solicited} \\ 
\end{tabular}\\
$^*$ -- This parameter was initially included in the model, but was later removed because it could not be reliably estimated. \\
\end{table}

\newpage
\subsection*{MCMC diagnosis}

\begin{table}[H]
\caption{\bf{MCMC performance for analyses of archived alignments or trees.}} \label{tab:mcmc_archive_any}
\centering
\begin{tabular}{lcccccccccc}
\toprule
& \multicolumn{5}{c}{Effective Sample Size} & \multicolumn{4}{c}{Geweke's Diagnostic ($p$-value)} & \\
\cmidrule(r){2-6} \cmidrule(r){7-10}
Parameter & run 1 & run 2 & run 3 & run 4 & combined & run 1 & run 2 & run 3 & run 4 & PSRF \\ 
\midrule
$\beta_\text{I}$ & 7501.00 & 7501.00 & 7501.00 & 7428.18 & 29931.18 & 0.36 & 0.07 & 0.72 & 0.64 & 1.00 \\ 
\rowcolor{gray!25}
$\beta_\text{age}$ & 7770.81 & 7501.00 & 7770.65 & 6917.10 & 29959.56 & 0.74 & 0.37 & 0.98 & 0.64 & 1.00 \\ 
$\beta_\text{IF}$ & 7236.37 & 7501.00 & 7501.00 & 7501.00 & 29739.37 & 0.59 & 0.47 & 0.11 & 0.06 & 1.00 \\ 
\rowcolor{gray!25}
$\beta_\text{none}$ & 7501.00 & 7501.00 & 8372.49 & 7489.66 & 30864.15 & 0.47 & 0.01 & 0.68 & 0.32 & 1.00 \\ 
$\beta_\text{strong}$ & 7501.00 & 7501.00 & 7220.13 & 7501.00 & 29723.13 & 0.75 & 0.55 & 0.57 & 0.54 & 1.00 \\ 
\rowcolor{gray!25}
$\beta_\text{JDAP}$ & 7501.00 & 7501.00 & 7501.00 & 6773.91 & 29276.91 & 0.16 & 0.18 & 0.83 & 0.84 & 1.00 \\ 
$\beta_\text{NSF}$ & 7501.00 & 7501.00 & 7964.70 & 7165.25 & 30131.94 & 0.71 & 0.61 & 0.55 & 0.08 & 1.00 \\ 
\end{tabular}
\begin{flushleft}N.B. This corresponds to analysis $x_{a,e}$ in Tables \ref{tab:modeltable} and \ref{tab:paramtable}.
\end{flushleft}
\end{table}

\begin{table}[H]
\caption{\bf{MCMC performance for analyses of archived alignments and trees.}} \label{tab:mcmc_archive_both}
\centering
\begin{tabular}{lcccccccccc}
\toprule
& \multicolumn{5}{c}{Effective Sample Size} & \multicolumn{4}{c}{Geweke's Diagnostic ($p$-value)} & \\
\cmidrule(r){2-6} \cmidrule(r){7-10}
Parameter & run 1 & run 2 & run 3 & run 4 & combined & run 1 & run 2 & run 3 & run 4 & PSRF \\ 
\midrule
$\beta_\text{I}$ & 7501.00 & 7501.00 & 7252.12 & 7501.00 & 29755.12 & 0.29 & 0.08 & 0.77 & 0.21 & 1.00 \\ 
\rowcolor{gray!25}
$\beta_\text{age}$ & 7501.00 & 7501.00 & 7501.00 & 7821.37 & 30324.37 & 0.26 & 0.11 & 0.14 & 0.86 & 1.00 \\ 
$\beta_\text{IF}$ & 7501.00 & 6147.95 & 7501.00 & 7501.00 & 28650.95 & 0.23 & 0.73 & 0.03 & 0.50 & 1.00 \\ 
\rowcolor{gray!25}
$\beta_\text{none}$ & 7501.00 & 7501.00 & 7501.00 & 7501.00 & 30004.00 & 0.15 & 0.04 & 0.29 & 0.95 & 1.00 \\ 
$\beta_\text{strong}$ & 7616.62 & 7501.00 & 7501.00 & 7501.00 & 30119.62 & 0.76 & 0.03 & 0.25 & 0.07 & 1.00 \\ 
\rowcolor{gray!25}
$\beta_\text{JDAP}$ & 7501.00 & 7501.00 & 7501.00 & 6887.44 & 29390.44 & 0.45 & 0.08 & 0.16 & 0.20 & 1.00 \\ 
$\beta_\text{NSF}$ & 7395.85 & 7501.00 & 7501.00 & 7501.00 & 29898.85 & 0.05 & 0.66 & 0.31 & 0.54 & 1.00 \\ 
\end{tabular}
\begin{flushleft}N.B. This corresponds to analysis $x_{a,b}$ in Tables \ref{tab:modeltable} and \ref{tab:paramtable}.
\end{flushleft}
\end{table}

\begin{table}[H]
\caption{\bf{MCMC performance for analyses of solicited alignments or trees.}} \label{tab:mcmc_sent_any}
\centering
\begin{tabular}{lcccccccccc}
\toprule
& \multicolumn{5}{c}{Effective Sample Size} & \multicolumn{4}{c}{Geweke's Diagnostic ($p$-value)} & \\
\cmidrule(r){2-6} \cmidrule(r){7-10}
Parameter & run 1 & run 2 & run 3 & run 4 & combined & run 1 & run 2 & run 3 & run 4 & PSRF \\ 
\midrule
$\beta_\text{I}$ & 7501.00 & 7501.00 & 7501.00 & 7501.00 & 30004.00 & 0.38 & 0.40 & 0.58 & 0.20 & 1.00 \\ 
\rowcolor{gray!25}
$\beta_\text{age}$ & 7501.00 & 7501.00 & 7501.00 & 7501.00 & 30004.00 & 0.48 & 0.57 & 0.68 & 0.51 & 1.00 \\ 
$\beta_\text{IF}$ & 7501.00 & 7501.00 & 7501.00 & 7501.00 & 30004.00 & 0.04 & 0.58 & 0.32 & 0.76 & 1.00 \\ 
\rowcolor{gray!25}
$\beta_\text{none}$ & 7501.00 & 7501.00 & 7501.00 & 7501.00 & 30004.00 & 0.24 & 0.28 & 0.43 & 0.77 & 1.00 \\ 
$\beta_\text{strong}$ & 7501.00 & 7501.00 & 7501.00 & 7501.00 & 30004.00 & 0.63 & 0.84 & 0.82 & 0.73 & 1.00 \\ 
\rowcolor{gray!25}
$\beta_\text{JDAP}$ & 7501.00 & 8065.92 & 7501.00 & 7501.00 & 30568.92 & 0.87 & 0.16 & 0.64 & 0.39 & 1.00 \\ 
$\beta_\text{NSF}$ & 7501.00 & 7501.00 & 7501.00 & 7383.98 & 29886.98 & 0.56 & 0.87 & 0.97 & 0.96 & 1.00 \\ 
\rowcolor{gray!25}
$\beta_\text{undergrad}$ & 7501.00 & 7009.73 & 7501.00 & 7501.00 & 29512.73 & 0.03 & 0.02 & 0.58 & 0.32 & 1.00 \\ 
$\beta_\text{prof}$ & 7501.00 & 7235.82 & 7501.00 & 8881.95 & 31119.77 & 0.74 & 0.68 & 0.51 & 0.04 & 1.00 \\ 
\end{tabular}
\begin{flushleft}N.B. This corresponds to analysis $x_{s,e}$ in Tables \ref{tab:modeltable} and \ref{tab:paramtable}.
\end{flushleft}
\end{table}

\begin{table}[H]
\caption{\bf{MCMC performance for analyses of solicited alignments and trees.}} \label{tab:mcmc_sent_both}
\centering
\begin{tabular}{lcccccccccc}
\toprule
& \multicolumn{5}{c}{Effective Sample Size} & \multicolumn{4}{c}{Geweke's Diagnostic ($p$-value)} & \\
\cmidrule(r){2-6} \cmidrule(r){7-10}
Parameter & run 1 & run 2 & run 3 & run 4 & combined & run 1 & run 2 & run 3 & run 4 & PSRF \\ 
\midrule
$\beta_\text{I}$ & 7501.00 & 7221.13 & 8541.62 & 7501.00 & 30764.75 & 0.84 & 0.33 & 0.29 & 0.90 & 1.00 \\ 
\rowcolor{gray!25}
$\beta_\text{age}$ & 7501.00 & 7223.68 & 7501.00 & 7501.00 & 29726.68 & 0.20 & 0.98 & 0.82 & 0.41 & 1.00 \\ 
$\beta_\text{IF}$ & 7501.00 & 7501.00 & 8580.70 & 7562.06 & 31144.76 & 0.90 & 0.92 & 0.37 & 0.31 & 1.00 \\ 
\rowcolor{gray!25}
$\beta_\text{none}$ & 7501.00 & 7501.00 & 7501.00 & 7501.00 & 30004.00 & 0.09 & 0.23 & 0.56 & 0.17 & 1.00 \\ 
$\beta_\text{strong}$ & 7501.00 & 7501.00 & 7501.00 & 7501.00 & 30004.00 & 1.00 & 0.77 & 0.62 & 0.58 & 1.00 \\ 
\rowcolor{gray!25}
$\beta_\text{JDAP}$ & 7501.00 & 7501.00 & 7501.00 & 7501.00 & 30004.00 & 0.18 & 0.20 & 0.69 & 0.14 & 1.00 \\ 
$\beta_\text{NSF}$ & 7501.00 & 7501.00 & 7501.00 & 7501.00 & 30004.00 & 0.97 & 0.23 & 0.74 & 0.10 & 1.00 \\ 
\rowcolor{gray!25}
$\beta_\text{undergrad}$ & 7185.12 & 7501.00 & 7501.00 & 7501.00 & 29688.12 & 0.17 & 0.93 & 0.40 & 0.86 & 1.00 \\ 
$\beta_\text{prof}$ & 7256.96 & 7501.00 & 7501.00 & 7501.00 & 29759.96 & 0.28 & 0.64 & 0.02 & 0.04 & 1.00 \\ 
\end{tabular}
\begin{flushleft}N.B. This corresponds to analysis $x_{s,b}$ in Tables \ref{tab:modeltable} and \ref{tab:paramtable}.
\end{flushleft}
\end{table}

\begin{table}[H]
\caption{\bf{MCMC performance for analyses of archived alignments only.}} \label{tab:mcmc_archive_alignments}
\centering
\begin{tabular}{lcccccccccc}
\toprule
& \multicolumn{5}{c}{Effective Sample Size}  & \multicolumn{4}{c}{Geweke's Diagnostic ($p$-value)} & \\
\cmidrule(r){2-6} \cmidrule(r){7-10}
Parameter & run 1 & run 2 & run 3 & run 4 & combined & run 1 & run 2 & run 3 & run 4 & PSRF \\ 
\midrule
$\beta_\text{I}$ & 7501.00 & 7501.00 & 7337.49 & 7830.03 & 30169.52 & 0.10 & 0.14 & 0.85 & 0.21 & 1.00 \\ 
\rowcolor{gray!25}
$\beta_\text{age}$ & 6993.23 & 7501.00 & 7501.00 & 7501.00 & 29496.23 & 0.24 & 0.85 & 0.92 & 0.62 & 1.00 \\ 
$\beta_\text{IF}$ & 8214.77 & 7501.00 & 7501.00 & 7501.00 & 30717.77 & 0.76 & 0.79 & 0.43 & 0.08 & 1.00 \\ 
\rowcolor{gray!25}
$\beta_\text{none}$ & 7501.00 & 7501.00 & 7501.00 & 7501.00 & 30004.00 & 0.56 & 0.18 & 0.74 & 0.73 & 1.00 \\ 
$\beta_\text{strong}$ & 7501.00 & 7501.00 & 7501.00 & 7501.00 & 30004.00 & 0.94 & 0.24 & 0.77 & 0.29 & 1.00 \\ 
\rowcolor{gray!25}
$\beta_\text{JDAP}$ & 7501.00 & 8255.15 & 7855.49 & 7501.00 & 31112.64 & 0.20 & 0.03 & 0.97 & 0.39 & 1.00 \\ 
$\beta_\text{NSF}$ & 7501.00 & 7501.00 & 7501.00 & 7501.00 & 30004.00 & 0.65 & 0.39 & 0.49 & 0.03 & 1.00 \\ 
\end{tabular}
\begin{flushleft}N.B. This corresponds to analysis $x_{a,a}$ in Tables \ref{tab:modeltable} and \ref{tab:paramtable}.
\end{flushleft}
\end{table}

\begin{table}[H]
\caption{\bf{MCMC performance for analyses of archived trees only.}} \label{tab:mcmc_archive_trees}
\centering
\begin{tabular}{lcccccccccc}
\toprule
& \multicolumn{5}{c}{Effective Sample Size} & \multicolumn{4}{c}{Geweke's Diagnostic ($p$-value)} & \\
\cmidrule(r){2-6} \cmidrule(r){7-10}
Parameter & run 1 & run 2 & run 3 & run 4 & combined & run 1 & run 2 & run 3 & run 4 & PSRF \\ 
\midrule
$\beta_\text{I}$ & 7501.00 & 7501.00 & 7501.00 & 7669.34 & 30172.34 & 0.66 & 0.46 & 0.67 & 0.44 & 1.00 \\ 
\rowcolor{gray!25}
$\beta_\text{age}$ & 7501.00 & 7092.89 & 7501.00 & 7501.00 & 29595.89 & 0.74 & 0.43 & 0.93 & 0.02 & 1.00 \\ 
$\beta_\text{IF}$ & 7501.00 & 7501.00 & 6660.60 & 7501.00 & 29163.60 & 0.83 & 0.47 & 0.85 & 0.68 & 1.00 \\ 
\rowcolor{gray!25}
$\beta_\text{none}$ & 7501.00 & 7501.00 & 7501.00 & 7501.00 & 30004.00 & 0.37 & 0.34 & 0.17 & 0.43 & 1.00 \\ 
$\beta_\text{strong}$ & 7501.00 & 7501.00 & 7501.00 & 7501.00 & 30004.00 & 0.42 & 0.73 & 0.68 & 0.34 & 1.00 \\ 
\rowcolor{gray!25}
$\beta_\text{JDAP}$ & 7501.00 & 7501.00 & 7501.00 & 7068.63 & 29571.63 & 0.96 & 0.63 & 0.33 & 0.66 & 1.00 \\ 
$\beta_\text{NSF}$ & 7501.00 & 7501.00 & 7501.00 & 7501.00 & 30004.00 & 0.61 & 0.82 & 0.63 & 0.26 & 1.00 \\ 
\end{tabular}
\begin{flushleft}N.B. This corresponds to analysis $x_{a,t}$ in Tables \ref{tab:modeltable} and \ref{tab:paramtable}.
\end{flushleft}
\end{table}

\begin{table}[H]
\caption{\bf{MCMC performance for analyses of solicited alignments only.}} \label{tab:mcmc_sent_alignment}
\centering
\begin{tabular}{lcccccccccc}
\toprule
& \multicolumn{5}{c}{Effective Sample Size} & \multicolumn{4}{c}{Geweke's Diagnostic ($p$-value)} & \\
\cmidrule(r){2-6} \cmidrule(r){7-10}
Parameter & run 1 & run 2 & run 3 & run 4 & combined & run 1 & run 2 & run 3 & run 4 & PSRF \\ 
\midrule
$\beta_\text{I}$ & 7501.00 & 7501.00 & 7501.00 & 7501.00 & 30004.00 & 0.95 & 0.31 & 0.85 & 0.98 & 1.00 \\ 
\rowcolor{gray!25}
$\beta_\text{age}$ & 7501.00 & 7501.00 & 7501.00 & 7501.00 & 30004.00 & 0.78 & 0.38 & 0.73 & 0.28 & 1.00 \\ 
$\beta_\text{IF}$ & 7752.87 & 7788.54 & 7501.00 & 7501.00 & 30543.41 & 0.81 & 0.65 & 0.98 & 0.81 & 1.00 \\ 
\rowcolor{gray!25}
$\beta_\text{none}$ & 7884.20 & 7501.00 & 7501.00 & 7501.00 & 30387.20 & 0.38 & 0.47 & 0.97 & 0.78 & 1.00 \\ 
$\beta_\text{strong}$ & 7501.00 & 7757.03 & 7501.00 & 8110.93 & 30869.96 & 0.55 & 0.35 & 0.62 & 0.66 & 1.00 \\ 
\rowcolor{gray!25}
$\beta_\text{undergrad}$ & 7501.00 & 7501.00 & 7501.00 & 8190.22 & 30693.22 & 0.77 & 0.27 & 0.67 & 0.73 & 1.00 \\ 
$\beta_\text{prof}$ & 7501.00 & 7501.00 & 7501.00 & 7857.31 & 30360.31 & 0.98 & 0.96 & 0.96 & 0.60 & 1.00 \\ 
\end{tabular}
\begin{flushleft}N.B. This corresponds to analysis $x_{s,a}$ in Tables \ref{tab:modeltable} and \ref{tab:paramtable}.
\end{flushleft}
\end{table}

\begin{table}[H]
\caption{\bf{MCMC performance for analyses of solicited trees only.}} \label{tab:mcmc_sent_trees}
\centering
\begin{tabular}{lcccccccccc}
\toprule
& \multicolumn{5}{c}{Effective Sample Size} & \multicolumn{4}{c}{Geweke's Diagnostic ($p$-value)} & \\
\cmidrule(r){2-6} \cmidrule(r){7-10}
Parameter & run 1 & run 2 & run 3 & run 4 & combined & run 1 & run 2 & run 3 & run 4 & PSRF \\ 
\midrule
$\beta_\text{I}$ & 7501.00 & 7501.00 & 7501.00 & 7501.00 & 30004.00 & 0.68 & 0.76 & 1.00 & 0.33 & 1.00 \\ 
\rowcolor{gray!25}
$\beta_\text{age}$ & 7501.00 & 7501.00 & 7501.00 & 7501.00 & 30004.00 & 0.34 & 0.47 & 0.92 & 0.83 & 1.00 \\ 
$\beta_\text{IF}$ & 7501.00 & 7501.00 & 7501.00 & 7501.00 & 30004.00 & 0.71 & 0.24 & 0.58 & 0.64 & 1.00 \\ 
\rowcolor{gray!25}
$\beta_\text{none}$ & 7229.02 & 7501.00 & 7501.00 & 7806.79 & 30037.81 & 0.46 & 0.34 & 0.66 & 0.29 & 1.00 \\ 
$\beta_\text{strong}$ & 7501.00 & 7501.00 & 7501.00 & 7501.00 & 30004.00 & 0.92 & 0.52 & 0.41 & 0.13 & 1.00 \\ 
\rowcolor{gray!25}
$\beta_\text{JDAP}$ & 7501.00 & 7501.00 & 7203.03 & 7501.00 & 29706.03 & 0.34 & 0.31 & 0.27 & 0.10 & 1.00 \\ 
$\beta_\text{NSF}$ & 7501.00 & 7501.00 & 7231.46 & 7501.00 & 29734.46 & 0.20 & 0.71 & 0.69 & 0.69 & 1.00 \\ 
\rowcolor{gray!25}
$\beta_\text{undergrad}$ & 7501.00 & 7501.00 & 7501.00 & 7501.00 & 30004.00 & 0.60 & 0.98 & 0.58 & 0.84 & 1.00 \\ 
$\beta_\text{prof}$ & 7429.18 & 7501.00 & 7501.00 & 7501.00 & 29932.18 & 0.52 & 0.97 & 0.72 & 0.38 & 1.00 \\ 
\end{tabular}
\begin{flushleft}N.B. This corresponds to analysis $x_{s,t}$ in Tables \ref{tab:modeltable} and \ref{tab:paramtable}.
\end{flushleft}
\end{table}

\begin{table}[H]
\caption{\bf{MCMC performance for analyses of combined alignments only.}} \label{tab:mcmc_combined_alignments}
\centering
\begin{tabular}{lcccccccccc}
\toprule
& \multicolumn{5}{c}{Effective Sample Size} & \multicolumn{4}{c}{Geweke's Diagnostic ($p$-value)} & \\
\cmidrule(r){2-6} \cmidrule(r){7-10}
Parameter & run 1 & run 2 & run 3 & run 4 & combined & run 1 & run 2 & run 3 & run 4 & PSRF \\ 
\midrule
$\beta_\text{I}$ & 8932.81 & 7501.00 & 7501.00 & 7501.00 & 31435.81 & 0.53 & 0.38 & 0.36 & 0.85 & 1.00 \\ 
\rowcolor{gray!25}
$\beta_\text{age}$ & 7501.00 & 7192.10 & 7501.00 & 7203.17 & 29397.27 & 0.39 & 0.61 & 0.07 & 0.36 & 1.00 \\ 
$\beta_\text{IF}$ & 7344.84 & 7501.00 & 7501.00 & 7501.00 & 29847.84 & 0.61 & 0.31 & 0.91 & 0.38 & 1.00 \\ 
\rowcolor{gray!25}
$\beta_\text{none}$ & 7501.00 & 7904.21 & 7501.00 & 7501.00 & 30407.21 & 0.47 & 0.30 & 0.24 & 0.11 & 1.00 \\ 
$\beta_\text{strong}$ & 8118.81 & 7122.63 & 7501.00 & 7795.53 & 30537.97 & 0.61 & 0.99 & 0.07 & 0.75 & 1.00 \\ 
\rowcolor{gray!25}
$\beta_\text{JDAP}$ & 7690.78 & 7501.00 & 7501.00 & 7501.00 & 30193.78 & 0.69 & 0.79 & 0.12 & 0.34 & 1.00 \\ 
$\beta_\text{NSF}$ & 7501.00 & 7501.00 & 7501.00 & 7501.00 & 30004.00 & 0.53 & 0.66 & 0.00 & 0.76 & 1.00 \\ 
\rowcolor{gray!25}
$\beta_\text{solicited}$ & 7501.00 & 7903.44 & 7501.00 & 7894.54 & 30799.98 & 0.72 & 0.10 & 0.05 & 0.05 & 1.00 \\ 
\end{tabular}
\begin{flushleft}N.B. This corresponds to analysis $x_{c,a}$ in Tables \ref{tab:modeltable} and \ref{tab:paramtable}.
\end{flushleft}
\end{table}

\begin{table}[H]
\caption{\bf{MCMC performance for analyses of combined trees only.}} \label{tab:mcmc_combined_trees}
\centering
\begin{tabular}{lcccccccccc}
\toprule
& \multicolumn{5}{c}{Effective Sample Size} & \multicolumn{4}{c}{Geweke's Diagnostic ($p$-value)} & \\
\cmidrule(r){2-6} \cmidrule(r){7-10}
Parameter & run 1 & run 2 & run 3 & run 4 & combined & run 1 & run 2 & run 3 & run 4 & PSRF \\ 
\midrule
$\beta_\text{I}$ & 7284.35 & 7501.00 & 7501.00 & 7012.23 & 29298.58 & 1.00 & 0.96 & 0.14 & 0.38 & 1.00 \\ 
\rowcolor{gray!25}
$\beta_\text{age}$ & 6638.97 & 7501.00 & 7501.00 & 7501.00 & 29141.97 & 0.97 & 0.26 & 0.02 & 0.69 & 1.00 \\ 
$\beta_\text{IF}$ & 7501.00 & 7501.00 & 7716.11 & 7401.50 & 30119.61 & 0.71 & 0.66 & 0.96 & 0.46 & 1.00 \\ 
\rowcolor{gray!25}
$\beta_\text{none}$ & 7501.00 & 7894.56 & 7501.00 & 7501.00 & 30397.56 & 0.74 & 0.36 & 0.98 & 0.81 & 1.00 \\ 
$\beta_\text{strong}$ & 7501.00 & 7501.00 & 6726.40 & 7501.00 & 29229.40 & 0.91 & 0.89 & 0.15 & 0.15 & 1.00 \\ 
\rowcolor{gray!25}
$\beta_\text{JDAP}$ & 7501.00 & 7501.00 & 8298.20 & 7233.11 & 30533.31 & 0.75 & 0.22 & 0.63 & 0.78 & 1.00 \\ 
$\beta_\text{NSF}$ & 7501.00 & 7501.00 & 7501.00 & 7501.00 & 30004.00 & 0.63 & 0.37 & 0.39 & 0.12 & 1.00 \\ 
\rowcolor{gray!25}
$\beta_\text{requested}$ & 7501.00 & 7501.00 & 7501.00 & 7501.00 & 30004.00 & 0.27 & 0.66 & 0.23 & 0.14 & 1.00 \\ 
\end{tabular}
\begin{flushleft}N.B. This corresponds to analysis $x_{c,t}$ in Tables \ref{tab:modeltable} and \ref{tab:paramtable}.
\end{flushleft}
\end{table}

\begin{table}[H]
\caption{\bf{MCMC performance for analyses of combined alignments or trees.}} \label{tab:mcmc_combined_any}
\centering
\begin{tabular}{lcccccccccc}
\toprule
& \multicolumn{5}{c}{Effective Sample Size} & \multicolumn{4}{c}{Geweke's Diagnostic ($p$-value)} & \\
\cmidrule(r){2-6} \cmidrule(r){7-10}
Parameter & run 1 & run 2 & run 3 & run 4 & combined & run 1 & run 2 & run 3 & run 4 & PSRF \\ 
\midrule
$\beta_\text{I}$ & 6261.86 & 7501.00 & 7501.00 & 6837.35 & 28101.21 & 0.25 & 0.42 & 0.85 & 0.45 & 1.00 \\ 
\rowcolor{gray!25}
$\beta_\text{age}$ & 7501.00 & 7501.00 & 7501.00 & 7501.00 & 30004.00 & 0.10 & 0.79 & 0.26 & 0.29 & 1.00 \\ 
$\beta_\text{IF}$ & 7501.00 & 7501.00 & 7501.00 & 7501.00 & 30004.00 & 0.90 & 0.43 & 0.27 & 0.76 & 1.00 \\ 
\rowcolor{gray!25}
$\beta_\text{none}$ & 7501.00 & 7501.00 & 7501.00 & 7501.00 & 30004.00 & 0.09 & 0.39 & 0.71 & 0.36 & 1.00 \\ 
$\beta_\text{strong}$ & 7501.00 & 7501.00 & 7501.00 & 7501.00 & 30004.00 & 0.00 & 0.18 & 0.67 & 0.78 & 1.00 \\ 
\rowcolor{gray!25}
$\beta_\text{JDAP}$ & 7521.48 & 7501.00 & 7501.00 & 7240.79 & 29764.27 & 0.08 & 0.57 & 0.13 & 0.40 & 1.00 \\ 
$\beta_\text{NSF}$ & 7730.56 & 7501.00 & 7501.00 & 7271.97 & 30004.53 & 0.34 & 0.85 & 0.40 & 0.76 & 1.00 \\ 
\rowcolor{gray!25}
$\beta_\text{requested}$ & 7501.00 & 7501.00 & 7501.00 & 7501.00 & 30004.00 & 0.11 & 0.74 & 0.35 & 0.85 & 1.00 \\ 
\end{tabular}
\begin{flushleft}N.B. This corresponds to analysis $x_{c,e}$ in Tables \ref{tab:modeltable} and \ref{tab:paramtable}.
\end{flushleft}
\end{table}

\begin{table}[H]
\caption{\bf{MCMC performance for analyses of combined alignments and trees.}} \label{tab:mcmc_combined_both}
\centering
\begin{tabular}{lcccccccccc}
\toprule
Parameter & \multicolumn{5}{c}{Effective Sample Size} & \multicolumn{4}{c}{Geweke's Diagnostic ($p$-value)} & \\
\cmidrule(r){2-6} \cmidrule(r){7-10}
& run 1 & run 2 & run 3 & run 4 & combined & run 1 & run 2 & run 3 & run 4 & PSRF \\ 
\midrule
$\beta_\text{I}$ & 7501.00 & 7501.00 & 7501.00 & 7501.00 & 30004.00 & 0.44 & 0.38 & 0.62 & 0.75 & 1.00 \\ 
\rowcolor{gray!25}
$\beta_\text{age}$ & 7501.00 & 7501.00 & 7501.00 & 7501.00 & 30004.00 & 0.51 & 0.09 & 0.46 & 0.22 & 1.00 \\ 
$\beta_\text{IF}$ & 7501.00 & 7501.00 & 7501.00 & 7501.00 & 30004.00 & 0.59 & 0.56 & 0.14 & 0.79 & 1.00 \\ 
\rowcolor{gray!25}
$\beta_\text{none}$ & 8623.45 & 7501.00 & 7501.00 & 7841.88 & 31467.33 & 0.77 & 0.72 & 0.44 & 0.91 & 1.00 \\ 
$\beta_\text{strong}$ & 7501.00 & 7501.00 & 7501.00 & 7501.00 & 30004.00 & 0.18 & 0.47 & 0.34 & 0.27 & 1.00 \\ 
\rowcolor{gray!25}
$\beta_\text{JDAP}$ & 7918.18 & 7501.00 & 7501.00 & 7281.23 & 30201.41 & 0.16 & 0.35 & 0.78 & 0.65 & 1.00 \\ 
$\beta_\text{NSF}$ & 7501.00 & 7501.00 & 7772.42 & 8128.34 & 30902.75 & 0.71 & 0.29 & 0.28 & 0.23 & 1.00 \\ 
\rowcolor{gray!25}
$\beta_\text{requested}$ & 7758.05 & 7501.00 & 7240.75 & 7501.00 & 30000.80 & 0.45 & 0.89 & 0.72 & 0.75 & 1.00 \\ 
\end{tabular}
\begin{flushleft}N.B. This corresponds to analysis $x_{c,a}$ in Tables \ref{tab:modeltable} and \ref{tab:paramtable}.
\end{flushleft}
\end{table}


\newpage
\subsection*{Logistic model parameter estimates}

\begin{table}[H]
\caption{\bf{Parameter estimates based on archived data.}}
\begin{tabular}{l ccc ccc ccc ccc}
\toprule
& \multicolumn{3}{c}{alignments only} & \multicolumn{3}{c}{trees only} & \multicolumn{3}{c}{alignments or trees} & \multicolumn{3}{c}{alignments and trees} \\
\cmidrule(r){2-4} \cmidrule(r){5-7} \cmidrule(r){8-10} \cmidrule(r){11-13}
& & \multicolumn{2}{c}{$95\%$ HPD} & & \multicolumn{2}{c}{$95\%$ HPD} & & \multicolumn{2}{c}{$95\%$ HPD} & & \multicolumn{2}{c}{$95\%$ HPD}  \\
\cmidrule(r){3-4} \cmidrule(r){6-7} \cmidrule(r){9-10} \cmidrule(r){12-13}
& mean & lower & upper & mean & lower & upper & mean & lower & upper & mean & lower & upper \\
\midrule
$\beta_\text{I}$ & -3.114 & -5.028 & -1.266 & -0.864 & -2.623 & 0.627 & -0.968 & -1.958 & 0.067 & -3.172 & -5.009 & -1.459 \\ 
\rowcolor{gray!25}
$\beta_\text{age}$ & -0.033 & -0.069 & 0.002 & -0.080 & -0.139 & -0.026 & -0.038 & -0.060 & -0.016 & -0.019 & -0.043 & 0.003 \\ 
$\beta_\text{IF}$ & -0.018 & -0.127 & 0.084 & 0.035 & -0.068 & 0.141 & 0.036 & -0.028 & 0.097 & 0.045 & -0.028 & 0.118 \\ 
\rowcolor{gray!25}
$\beta_\text{none}$ & 0.650 & -1.357 & 2.761 & -0.627 & -2.615 & 1.310 & 0.234 & -0.886 & 1.386 & 0.783 & -1.274 & 2.883 \\ 
$\beta_\text{strong}$ & 1.001 & -1.023 & 3.293 & -0.082 & -2.323 & 1.903 & 1.075 & -0.044 & 2.284 & 1.722 & -0.087 & 3.712 \\ 
\rowcolor{gray!25}
$\beta_\text{JDAP}$ & 1.685 & -0.114 & 3.594 & 0.066 & -1.529 & 1.791 & 2.370 & 1.382 & 3.528 & 3.048 & 1.289 & 4.854 \\ 
$\beta_\text{NSF}$ & 1.183 & -0.045 & 2.446 & -0.259 & -1.769 & 1.290 & 0.498 & -0.355 & 1.393 & -0.143 & -1.234 & 0.928 \\ 
\label{tab:results_archive_alignments}
\end{tabular}
\end{table}

\begin{table}[H]
\caption{\bf{Parameter estimates based on solicited data.}}
\centering
\begin{tabular}{l ccc ccc ccc ccc}
\toprule
& \multicolumn{3}{c}{alignments only} & \multicolumn{3}{c}{trees only} & \multicolumn{3}{c}{alignments or trees} & \multicolumn{3}{c}{alignments and trees} \\
\cmidrule(r){2-4} \cmidrule(r){5-7} \cmidrule(r){8-10} \cmidrule(r){11-13}
& & \multicolumn{2}{c}{$95\%$ HPD} & & \multicolumn{2}{c}{$95\%$ HPD} & & \multicolumn{2}{c}{$95\%$ HPD} & & \multicolumn{2}{c}{$95\%$ HPD}  \\
\cmidrule(r){3-4} \cmidrule(r){6-7} \cmidrule(r){9-10} \cmidrule(r){12-13}
& mean & lower & upper & mean & lower & upper & mean & lower & upper & mean & lower & upper \\
\midrule
$\beta_\text{I}$ & -6.248 & -9.587 & -3.130 & -1.577 & -3.096 & -0.248 & -0.328 & -1.225 & 0.548 & -1.708 & -2.780 & -0.619 \\ 
\rowcolor{gray!25}
$\beta_\text{age}$ & 0.004 & -0.018 & 0.027 & 0.001 & -0.018 & 0.022 & -0.004 & -0.016 & 0.007 & -0.006 & -0.019 & 0.005 \\ 
$\beta_\text{IF}$ & -0.338 & -0.733 & 0.024 & 0.130 & 0.034 & 0.222 & 0.081 & -0.003 & 0.171 & 0.002 & -0.076 & 0.075 \\ 
\rowcolor{gray!25}
$\beta_\text{none}$ & 0.653 & -1.530 & 2.787 & 0.578 & -0.953 & 2.062 & 0.281 & -0.599 & 1.127 & -0.089 & -1.013 & 0.805 \\ 
$\beta_\text{strong}$ & 1.555 & -0.622 & 3.810 & 0.209 & -1.466 & 1.876 & 0.573 & -0.425 & 1.499 & 0.311 & -0.661 & 1.303 \\ 
\rowcolor{gray!25}
$\beta_\text{JDAP}$ & NA & NA & NA & 0.622 & -1.617 & 2.863 & 0.027 & -1.424 & 1.347 & -0.095 & -1.609 & 1.469 \\ 
$\beta_\text{NSF}$ & NA & NA & NA & 0.220 & -1.538 & 1.907 & -0.482 & -1.590 & 0.663 & -0.612 & -1.901 & 0.556 \\ 
\rowcolor{gray!25}
$\beta_\text{undergrad}$ & 2.177 & -0.388 & 4.978 & -2.736 & -4.622 & -0.960 & 0.478 & -0.356 & 1.231 & 1.538 & 0.602 & 2.598 \\ 
$\beta_\text{prof}$ & 1.507 & -1.153 & 4.508 & -2.334 & -3.991 & -0.819 & 1.501 & 0.653 & 2.378 & 2.598 & 1.649 & 3.645 \\ 
\label{tab:results_sent_alignments}
\end{tabular}
\end{table}

\begin{table}[H]
\caption{\bf{Parameter estimates based on archived and solicited data.}}
\centering
\begin{tabular}{l ccc ccc ccc ccc}
\toprule
& \multicolumn{3}{c}{alignments only} & \multicolumn{3}{c}{trees only} & \multicolumn{3}{c}{alignments or trees} & \multicolumn{3}{c}{alignments and trees} \\
\cmidrule(r){2-4} \cmidrule(r){5-7} \cmidrule(r){8-10} \cmidrule(r){11-13}
& & \multicolumn{2}{c}{$95\%$ HPD} & & \multicolumn{2}{c}{$95\%$ HPD} & & \multicolumn{2}{c}{$95\%$ HPD} & & \multicolumn{2}{c}{$95\%$ HPD}  \\
\cmidrule(r){3-4} \cmidrule(r){6-7} \cmidrule(r){9-10} \cmidrule(r){12-13}
& mean & lower & upper & mean & lower & upper & mean & lower & upper & mean & lower & upper \\
\midrule
$\beta_\text{I}$ & -0.994 & -1.887 & -0.082 & 0.479 & -1.453 & 2.380 & 2.111 & 1.063 & 3.180 & -0.171 & -1.389 & 0.953 \\ 
\rowcolor{gray!25}
$\beta_\text{age}$ & -0.008 & -0.019 & 0.002 & -0.038 & -0.078 & 0.002 & -0.007 & -0.018 & 0.002 & 0.006 & -0.009 & 0.021 \\ 
$\beta_\text{IF}$ & -0.005 & -0.060 & 0.049 & 0.004 & -0.102 & 0.111 & 0.002 & -0.054 & 0.059 & -0.012 & -0.083 & 0.058 \\ 
\rowcolor{gray!25}
$\beta_\text{none}$ & -0.114 & -0.867 & 0.679 & -0.335 & -2.227 & 1.589 & 0.101 & -0.688 & 0.834 & 0.772 & -0.445 & 1.992 \\ 
$\beta_\text{strong}$ & 0.289 & -0.548 & 1.084 & -0.609 & -2.517 & 1.313 & 0.338 & -0.477 & 1.171 & 0.353 & -0.814 & 1.611 \\ 
\rowcolor{gray!25}
$\beta_\text{JDAP}$ & -0.033 & -1.066 & 0.916 & -1.334 & -3.174 & 0.506 & 0.431 & -0.656 & 1.513 & 1.238 & 0.010 & 2.467 \\ 
$\beta_\text{NSF}$ & 0.232 & -0.593 & 1.074 & -0.858 & -2.429 & 0.663 & -0.342 & -1.235 & 0.520 & 0.172 & -0.880 & 1.211 \\ 
\rowcolor{gray!25}
$\beta_\text{requested}$ & 0.690 & -0.098 & 1.438 & -3.634 & -5.544 & -1.835 & -2.040 & -2.948 & -1.132 & -3.162 & -4.099 & -2.239 \\ 
\label{tab:results_combined_alignments}
\end{tabular}
\end{table}


\newpage
\subsection*{Journal Policies}

\begin{table}[!ht]
\caption{\bf{Summary of journal policies.}}
\centering
\resizebox{\textwidth}{!}{%
\begin{tabular}{lll}
\toprule
Journal & Policy rating & Number of studies \\
\midrule
The American Naturalist & \emph{JDAP membership} & 5 \\ 
\rowcolor{gray!25}
Annals of Botany & \emph{weak policy} & 1 \\ 
Australian Systematic Botany & \emph{weak policy} & 2 \\ 
\rowcolor{gray!25}
Biological Journal of the Linnean Society$^\dagger$ & \emph{weak policy} & 4 \\ 
Biogeography & \emph{weak policy} & 15 \\ 
\rowcolor{gray!25}
BMC Biology & \emph{weak policy} & 3 \\ 
BMC Evolutionary Biology$^\dagger$ & \emph{no policy} & 12 \\ 
\rowcolor{gray!25}
BMC Plant Biology & \emph{weak policy} & 1 \\ 
Cladistics & \emph{weak policy} & 1 \\ 
\rowcolor{gray!25}
Copeia & \emph{weak policy} & 1 \\ 
Ecology$^\dagger$ & \emph{no policy} & 2 \\ 
\rowcolor{gray!25}
Ecology and Evolution & \emph{strong policy} & 2 \\ 
Ecology Letters & \emph{no policy} & 4 \\ 
\rowcolor{gray!25}
Evolution & \emph{JDAP membership} & 35 \\ 
Evolutionary Biology & \emph{weak policy} & 1 \\ 
\rowcolor{gray!25}
Hydrobiologia & \emph{no policy} & 1 \\ 
International Journal of Plant Sciences & \emph{no policy} & 1 \\ 
\rowcolor{gray!25}
International Journal of Microbial Ecology & \emph{strong policy} & 1 \\ 
Italian Journal of Zoology & \emph{no policy} & 1 \\ 
\rowcolor{gray!25}
Journal of Arid Environments & \emph{no policy} & 1 \\ 
Journal of Evolutionary Biology & \emph{JDAP membership} & 9 \\ 
\rowcolor{gray!25}
Journal of Heredity$^\dagger$ & \emph{strong policy} & 1 \\ 
Journal of Mammalogy & \emph{weak policy} & 1 \\ 
\rowcolor{gray!25}
Journal of Mammalian Evolution & \emph{no policy} & 1 \\ 
Journal of Theoretical Biology & \emph{no policy} & 1 \\ 
\rowcolor{gray!25}
Methods in Ecology and Evolution & \emph{strong policy} & 1 \\ 
Molecular Biology and Evolution$^\dagger$ & \emph{weak policy} & 2 \\ 
\rowcolor{gray!25}
Molecular Ecology & \emph{JDAP membership} & 8 \\ 
Molecular Phylogenetics and Evolution & \emph{no policy} & 31 \\ 
\rowcolor{gray!25}
Nature$^\dagger$ & \emph{weak policy} & 2 \\ 
New Phytologist & \emph{weak policy} & 1 \\ 
\rowcolor{gray!25}
PLOS ONE$^\dagger$ & \emph{weak policy} & 12 \\ 
PLOS Biology$^\dagger$ & \emph{weak policy} & 2 \\ 
\rowcolor{gray!25}
PeerJ & NA & 1 \\ 
Plant Systematics and Evolution & \emph{weak policy} & 2 \\ 
\rowcolor{gray!25}
PNAS & \emph{weak policy} & 11 \\ 
Perspectives in Plant Ecology, Evolution and Systematics & \emph{no policy} & 1 \\ 
\rowcolor{gray!25}
Proceedings of the Royal Society B & \emph{strong policy} & 14 \\ 
Science & \emph{strong policy} & 5 \\ 
\rowcolor{gray!25}
Systematic Entomology & \emph{no policy} & 2 \\ 
Systematic Biology & \emph{JDAP membership} & 13 \\ 
\rowcolor{gray!25}
Trends in Microbiology & \emph{no policy} & 1 \\ 
Virus Research & \emph{no policy} & 1 \\ 
\label{tab:journal_policies}
\end{tabular}}
\begin{flushleft} $^\dagger$Data-sharing policy in place at the time of study publication; has since become \emph{JDAP membership.}
\end{flushleft}
\end{table}

\bigskip \noindent \textbf{\emph{The American Naturalist} (strong policy, JDAP member)}

\noindent
The American Naturalist \emph{requires authors to deposit the data associated with accepted papers in a public archive}. For gene sequence data and phylogenetic trees, deposition in GenBank or TreeBASE, respectively, is required. There are many possible archives that may suit a particular data set, including the Dryad repository for ecological and evolutionary biology data (http://datadryad.org). All accession numbers for GenBank, TreeBASE, and Dryad must be included in accepted manuscripts before they go to Production. If the data are deposited somewhere else, please provide a link. If the data are culled from published literature, please deposit the collated data in Dryad for the convenience of your readers. Any impediments to data sharing should be brought to the attention of the editors at the time of submission so that appropriate arrangements can be worked out. For more, see the editorial on data.

\bigskip \noindent \textbf{\emph{Annals of Botany} (weak policy)}

\noindent
Before novel sequences for proteins or nucleotides can be published, authors are required to deposit their data with one of the principal databases comprising the International Nucleotide Sequence Database Collaboration: EMBL Nucleotide Sequence Database, GenBank, or the DNA Data Bank of Japan and to include an accession number in the paper.

\emph{Sequence matrices should only be included if alignment information is critical to the message of the paper}. Such matrices can be in colour but should not occupy more than one printed page. Larger matrices will only be printed by special agreement but may more readily be published electronically as Supplementary Information.

\bigskip \noindent \textbf{\emph{Australian Systematic Botany} (weak policy)}

\noindent
For all papers, whether presenting morphological, cytological or molecular data, voucher specimens must be cited, along with the herbarium where lodged.

\emph{All sequences used as data must be deposited in one of the international nucleotide sequence databases}, preferably GenBank, National Center for Biotechnology Information, 8600 Rockville Pike, Bethesda, MD 20894, USA. Email: gb-sub@ncbi.nlm.nih.gov. Request information at gsdb@gsdb.ncgr.org. Post-review final manuscript will not be accepted until sequence database accession numbers are included.

\bigskip \noindent \textbf{\emph{Biological Journal of the Linnean Society} (weak policy$^\dagger$)}

\noindent
\emph{Data that are integral to the paper must be made available in such a way as to enable readers to replicate, verify and build upon the conclusions published in the paper.} Any restriction on the availability of these data must be disclosed at the time of submission. Data may be included as part of the main article where practical. We \emph{recommend that data for which public repositories are widely used, and are accessible to all, should be deposited in such a repository prior to publication.} The appropriate linking details and identifier(s) should then be included in the publication and where possible the repository, to facilitate linking between the journal article and the data. If such a repository does not exist, data should be included as supporting information to the published paper or authors should agree to make their data available upon reasonable request.

\noindent 
NB: $^\dagger$Data-sharing policy in place at the time of study publication, which has since become \emph{JDAP membership.}

\bigskip \noindent \textbf{\emph{BMC Biology} (weak policy)}

\noindent
Submission of a manuscript to a BioMed Central journal implies that \emph{readily reproducible materials described in the manuscript, including all relevant raw data, will be freely available to any scientist wishing to use them} for non-commercial purposes.

Through a special arrangement with LabArchives, LLC, authors submitting manuscripts to BMC Biology can obtain a complimentary subscription to LabArchives with an allotment of 100MB of storage. LabArchives is an Electronic Laboratory Notebook which will enable scientists to share and publish data files in situ; you can then link your paper to these data. Data files linked to published articles are assigned digital object identifiers (DOIs) and will remain available in perpetuity. Use of LabArchives or similar data publishing services does not replace preexisting data deposition requirements, such as for nucleic acid sequences, protein sequences and atomic coordinates.

\emph{The Accession Numbers of any nucleic acid sequences, protein sequences or atomic coordinates cited in the manuscript should be provided,} in square brackets and include the corresponding database name

\bigskip \noindent \textbf{\emph{BMC Evolutionary Biology} (no policy$^\dagger$)}

\noindent
BMC Evolutionary Biology \emph{encourages} authors to deposit the data set(s) supporting the results reported in submitted manuscripts in a publicly-accessible data repository, when it is not possible to publish them as additional files. This section should only be included when supporting data are available and must include the name of the repository and the permanent identifier or accession number and persistent hyperlink(s) for the data set(s).

\noindent 
NB: $^\dagger$Data-sharing policy in place at the time of study publication, which has since become \emph{JDAP membership.}

\bigskip \noindent \textbf{\emph{BMC Plant Biology} (weak policy)}

\noindent
\emph{The Accession Numbers of any nucleic acid sequences, protein sequences or atomic coordinates cited in the manuscript should be provided}, in square brackets and include the corresponding database name$\cdots$

The databases for which we can provide direct links are: EMBL Nucleotide Sequence Database (EMBL), DNA Data Bank of Japan (DDBJ), GenBank at the NCBI (GenBank), Protein Data Bank (PDB), Protein Information Resource (PIR) and the Swiss-Prot Protein Database (Swiss-Prot).

\emph{BMC Plant Biology encourages authors to deposit the data set(s) supporting the results} reported in submitted manuscripts \emph{in a publicly-accessible data repository,} when it is not possible to publish them as additional files. This section should only be included when supporting data are available and must include the name of the repository and the permanent identifier or accession number and persistent hyperlink(s) for the data set(s).

Through a special arrangement with LabArchives, LLC, authors submitting manuscripts to BMC Plant Biology can obtain a complimentary subscription to LabArchives with an allotment of 100MB of storage. LabArchives is an Electronic Laboratory Notebook which will enable scientists to share and publish data files in situ; you can then link your paper to these data. Data files linked to published articles are assigned digital object identifiers (DOIs) and will remain available in perpetuity. Use of LabArchives or similar data publishing services does not replace preexisting data deposition requirements, such as for nucleic acid sequences, protein sequences and atomic coordinates.

\bigskip \noindent \textbf{\emph{Cladistics} (weak policy)}

\noindent
\emph{Cladistics requests the deposition of data matrices} and other material electronically \emph{for publication on the Willi Hennig Society Journal web site.} Please submit these data as e-mail attachments to the Associate Editor Mark Siddall (siddall@amnh.org) after receiving the tracking number for your manuscript from the Editor. These data will be made available to the referees but not to the community at large until such time as the paper is accepted. If the paper is not found to be acceptable for publication, the data and associated files will be removed from the directory structure and destroyed.

\bigskip \noindent \textbf{\emph{Copeia} (weak policy)}

\noindent
\emph{Analyses based on molecular sequence data must cite the relevant GenBank accession numbers in the text.}

\bigskip \noindent \textbf{\emph{Ecology} (no policy$^\dagger$)}

\noindent
The editors and publisher expect authors to make the data underlying published articles available. Although \emph{public data availability is not strictly a requirement for manuscripts published in Ecology}, Ecological Applications and Ecosphere at this time, any information on materials, methods or data necessary to verify the conclusions of the research reported must be made available to the Subject-matter Editor upon request.

\noindent 
NB: $^\dagger$Data-sharing policy in place at the time of study publication, which has since become \emph{JDAP membership.}

\bigskip \noindent \textbf{\emph{Ecology and Evolution} (strong policy)}

\noindent
The Journal Ecology and Evolution \emph{requires, as a condition for publication, that data supporting the results in the paper should be archived in an appropriate public archive,} such as GenBank, TreeBASE, Dryad, the Knowledge Network for Biocomplexity or other suitable long-term and stable public repositories. Data are important products of the scientific enterprise, and they should be preserved and usable for decades in the future. 

\bigskip \noindent \textbf{\emph{Ecology Letters} (no policy)}

\noindent
It is \emph{recommended that authors deposit the data supporting the results} in the paper in a publically accessible archive, such as Dryad (DataDryad.Org). Data are important products of scientific enterprise, and they should be preserved and remain usable in future decades. DNA sequences published in Ecology Letters should be deposited in the EMBL/GenBank/DDJB Nucleotide Sequence Databases. An accession number for each sequence must be included in the manuscript.

\bigskip \noindent \textbf{\emph{Evolution} (strong policy, JDAP membership)}

\noindent
 As a condition for publication, Evolution \emph{requires that data used in the paper are archived. DNA sequence data must be submitted to GenBank and phylogenetic data to TreeBASE. Other types of data must be deposited in an appropriate public archive} such as Dryad, the NCEAS Data Repository, or as supplementary online material associated with the paper published in Evolution. \emph{The data should be given with sufficient detail that, together with the contents of the paper, they allow each result in the published paper to be recreated.} Authors may elect to have the data publicly available at time of publication, or, if the technology of the archive allows, may opt to embargo access to the data for a period up to a year after publication. Exceptions may be granted at the discretion of the Editor-in-Chief, especially for sensitive information such as the location of endangered species. Authors must state their intention to archive their data when they submit their manuscript and must confirm that this has been done before the manuscript is sent to press. If a repository is to be cited, the citation should include the sequence name and accession number, if available. The basic format for citing electronic resources is: Author's Last Name, First initial. Title of data package (e.g., Data from “Article name”). Data Repository Name, Data identifier (or DOI), address/URL. Please include on your title page the location of your data or where you intend to archive your data.

\bigskip \noindent \textbf{\emph{Evolutionary Bioinformatics Online} (weak policy)}

\noindent
Authors publishing in Libertas Academica journals must agree to \emph{make freely available to other academic researchers any of the cells, clones of cells, DNA, antibodies, or other material used in the research reported and not available from commercial suppliers.}

Use DNA Databank of Japan, European Molecular Biology Laboratory, or GenBank.

Please \emph{ensure that accession numbers of any nucleaic acid sequences, protein sequences or atomic coordinates cited in the manuscript are provided} in square brackets with the corresponding database name.

Generally it is possible to provide direct links to data hosted on these databases: EMBL Nucleotide Sequence Database, DNA Data Bank of Japan, GenBank at the NCBI, Protein Data Bank, Protein Information Resource and the SwissProt Protein Database.

\bigskip \noindent \textbf{\emph{Evolutionary Biology} (weak policy)}

\noindent
\emph{DNA sequences must be submitted to GenBank} (NCBI - National Center for Biotechnology Information, Bethesda, USA) \emph{or to the EMBL Nucleotide Sequence Data Base} (EBI - European Institute of Bioinformatics, Cambridge, UK) and \emph{accession numbers must be provided} when the paper is accepted.

\emph{Authors are encouraged to submit other data types to online data sharing resources if such resources are available.}

\bigskip \noindent \textbf{\emph{Hydrobiologia} (no policy)}

\bigskip \noindent \textbf{\emph{International Journal of Plant Sciences} (no policy)}

\noindent
\emph{Material that is not integral to the body of the article and that substantially lengthens the print version of the article} (e.g., genetic and character matrices, \emph{extended cladograms}, extended tables) \emph{should be desgignated as appendixes} (table A1, etc.) and thus appear in the electronic edition of IJPS only. Exception: \emph{If voucher material is presented in your manuscript, this should be listed in the first appendix} (appendix A1), and this will appear in the print version of IJPS. To prepare your accession data, provide an appendix title and a sentence-style row of headings for the data. For each taxon sampled, include specimen voucher information and/or gene accession numbers, separated by commas. To save space, the taxa should be run together in a paragraph.

\bigskip \noindent \textbf{\emph{ISME Journal} (strong policy)}

\noindent
The ISME Journal will \emph{only review and publish manuscripts if the authors agree to make all data that cannot be published in the journal itself} (e.g. novel nucleotide sequences, structural data, or data from large-scale gene expression experiments) \emph{freely available in one of the public databases} (see Submission to public databases below). \emph{Accession codes must be provided} at the time a revised manuscript is returned to the Editorial Office. To avoid delays in publication of the manuscript, \emph{we encourage authors to deposit relevant data in public databases prior to submission.} 

\bigskip \noindent \textbf{\emph{Italian Journal of Zoology} (no policy)}

\bigskip \noindent \textbf{\emph{Journal of Arid Environments} (no policy)}

\noindent
Elsevier \emph{accepts electronic supplementary material to support and enhance} your scientific research. Supplementary files offer the author additional possibilities to publish supporting applications, high-resolution images, background datasets, sound clips and more. Supplementary files supplied will be published online alongside the electronic version of your article in Elsevier Web products, including ScienceDirect.

Electronic archiving of supplementary data enables readers to replicate, verify and build upon the conclusions published in your paper. We \emph{recommend that data should be deposited} in the data library PANGAEA 

\bigskip \noindent \textbf{\emph{Journal of Biogeography} (weak policy)}

\noindent
Consistent with widely adopted conventions in the field, it is a condition of publication that \emph{papers using new molecular sequences must place the sequences in an appropriate database} (e.g. GenBank). Relevant accession numbers should be provided in the final manuscript. \emph{Accession numbers are required for all sequences used in analyses,} including existing sequences in databases. Museum voucher numbers should also be provided where doing so constitutes the appropriate best practice and/or where this information could be of real value to future researchers. More generally, the journal recognizes that what is considered appropriate best practice regarding data publication/deposition may vary depending on factors such as the nature of the data, the funding sources involved, complexities of prior intellectual ownership issues, etc.  We therefore \emph{strongly encourage (where appropriate) but do not require (where it may not be) authors to publish/deposit data sets} in conjunction with papers being published in this journal.

Authors who wish to provide a consolidated statement of how other readers can access the data used in their paper \emph{may wish to refer to outside data repositories} where they have deposited their data, e.g. Dryad, Pangaea, or others. If so, this statement should be included after the Supporting Information section and before the Biosketch entry.

\emph{Additional materials and results (including supporting tables and figures) that are necessary but do not need to be included in the main paper must be compiled into Appendices}, which will be provided to readers as online Supporting Information. No more than three supplementary appendices are permitted (labelled Appendix S1 to Appendix S3). 

\bigskip \noindent \textbf{\emph{Journal of Evolutionary Biology} (strong policy, JDAP membership)}

\noindent
Submission of a manuscript to a BioMed Central journal implies that readily reproducible materials described in the manuscript, including \emph{all relevant raw data, will be freely available to any scientist wishing to use them for non-commercial purposes.}

The Journal of Evolutionary Biology requires, as a condition for publication, that \emph{data supporting the results in the paper should be archived in an appropriate public archive,} such as GenBank, TreeBASE, Dryad, the Knowledge Network for Biocomplexity or other suitable long-term and stable public repositories. Data are important products of the scientific enterprise, and they should be preserved and usable for decades in the future. 

All accepted papers should \emph{provide accession numbers or DOI} for data underlying the work that have been deposited, so that these can appear in the final accepted article.

\bigskip \noindent \textbf{\emph{Journal of Heredity} (strong policy$^\dagger$)}

\noindent
The primary data underlying the conclusions of an article are critical to the verifiability and transparency of the scientific enterprise, and \emph{should be preserved in usable form for decades in the future.} For this reason, Journal of Heredity has \emph{adopted the Joint Data Archiving Policy} (JDAP)(See Editorial, J Hered (2013) 104 (1):1. doi: 10.1093/jhered/ess137). \emph{The public archiving of all primary data is a requirement} of publication in Journal of Heredity. For data other than nucleotide and protein sequences submitted to GenBank, \emph{suitable archives include Dryad, TreeBASE} or the Knowledge Network for Biocomplexity. \emph{For accepted articles, JHered covers the cost to archive the data in Dryad.}

Authors may elect to have the data publicly available at time of publication, or may opt to embargo access to the data for a period up to a year after publication. Exceptions may be granted at the discretion of the Editor, especially for sensitive information, such as the location of endangered species.

\noindent 
NB: $^\dagger$Data-sharing policy in place at the time of study publication, which has since become \emph{JDAP membership.}

\bigskip \noindent \textbf{\emph{Journal of Mammalian Evolution} (no policy)}

\noindent
Authors submitting analyses of gene or amino acid sequences should be prepared to supply electronic versions of their sequence alignments \emph{if requested by reviewers.} These are for confidential examination by reviewers only, and reviewers are requested not to use these alignments for any purpose other than the review process.

If the reported research includes examination of \emph{voucher specimens (including tissues), the Museum catalogue number or tissue number if the former is not available must be provided.} If sequences from Genbank are used, the \emph{Museum catalogue number} listed on Genbank must be listed, as well as the Genbank accession number. Genbank accession numbers can be used alone, only in the event that the Museum catalogue number is not available through the Genbank record.

\bigskip \noindent \textbf{\emph{Journal of Mammology} (weak policy)}

\noindent
\emph{All DNA sequences must be submitted to GenBank, and accession numbers provided in the manuscript before publication.}

Supplemental ﬁles will be posted online-only and provides information that adds depth to a manuscript but is not essential to a reader’s understanding of the research (e.g., spreadsheets, databases, equations, video or audio ﬁles, tables and/or ﬁgures).

\bigskip \noindent \textbf{\emph{Journal of Theoretical Biology} (no policy)}

\noindent
Elsevier \emph{encourages authors to connect articles with external databases}, giving their readers one-click access to relevant databases that help to build a better understanding of the described research. Please refer to relevant database identifiers using the following format in your article

Elsevier \emph{accepts electronic supplementary material} to support and enhance your scientific research. Supplementary files offer the author additional possibilities to publish supporting applications, high-resolution images, background datasets, sound clips and more. Supplementary files supplied will be published online alongside the electronic version of your article in Elsevier Web products, including ScienceDirect

\bigskip \noindent \textbf{\emph{Methods in Ecology and Evolution} (strong policy)}

\noindent
Data are important products of the scientific enterprise, and they should be preserved and usable for decades in the future. The British Ecological Society thus \emph{expects that data} (or, for theoretical papers, mathematical and computer models) \emph{supporting the results in Methods in Ecology and Evolution papers will be archived in an appropriate public archive, such as Dryad, TreeBASE,} NERC data centre, GenBank, or another archive of the author's choice that provides comparable access and guarantee of preservation.

\emph{Sufficient details should be archived} so that a third party can reasonably interpret those data correctly, \emph{to allow each result in the published paper to be recreated and the analyses reported in the paper to be replicated} to support the conclusions made. \emph{Authors are welcome to archive more than this, but not less.}

If data have been previously archived then they should not be archived again. \emph{The original archive DOI or reference should be used as the source of the data.}

\bigskip \noindent \textbf{\emph{Molecular Biology and Evolution} (weak policy$^\dagger$)}

\noindent
Among the \emph{requirements for publication} in MBE is that \emph{authors make publicly available, free of charge, any alignment data}, strains, cell lines, or clones used in reported experiments, computer code essential to the analysis, \emph{and any other material or information necessary for the assessment and verification of findings} or interpretations presented in the publication.

\emph{Newly reported sequences must be deposited} in the DDBJ/EMBL/GenBank database (see below). \emph{Accession numbers must be included in the final version of the manuscript} and cannot be added at the proof stage. 

Newly reported nucleic acid and amino acid sequences, microarray data, structural coordinates, and all other essential information must be submitted to appropriate public databases (e.g., GenBank; the EMBL Nucleotide Sequence Database; DNA Database of Japan; the Protein Data Bank; Swiss-Prot; GEO; and Array-Express). 

\emph{When appropriate, material such as sequence alignments} and large tables \emph{can be published online as supplementary material} permanently linked to an article in the online journal. 

\noindent 
NB: $^\dagger$Data-sharing policy in place at the time of study publication, which has since become \emph{JDAP membership.}

\bigskip \noindent \textbf{\emph{Molecular Ecology} (strong policy, JDAP membership)}

\noindent
Molecular Ecology expects that \emph{data supporting the results in the paper should be archived in an appropriate public archive,} such as \emph{GenBank}, Gene Expression Omnibus, \emph{TreeBASE, Dryad,} the Knowledge Network for Biocomplexity, your own institutional or funder repository, or as Supporting Information on the Molecular Ecology web site. Data are important products of the scientific enterprise, and they should be preserved and usable for decades in the future. 

\emph{Authors are expected to archive the data supporting their results and conclusions along with sufficient details so that a third party can interpret them correctly.}

To enable readers to locate archived data from Molecular Ecology papers, \emph{we require that authors include a ‘Data Accessibility’ section} after the references (see below for details). This section must be present at initial submission.

\bigskip \noindent \textbf{\emph{Molecular Phylogenetics and Evolution} (no policy)}

\noindent
Elsevier \emph{encourages authors to connect articles with external databases,} giving their readers one-click access to relevant databases that help to build a better understanding of the described research.

Elsevier \emph{accepts} electronic supplementary \emph{material to support and enhance} your scientific research. Supplementary files offer the author additional possibilities to publish supporting applications, high-resolution images, background datasets, sound clips and more. Supplementary files supplied will be published online alongside the electronic version of your article in Elsevier Web products, including ScienceDirect: http://www.sciencedirect.com.

You can \emph{enrich} your online articles \emph{by providing phylogenetic tree data files (optional)} in Newick or NeXML format, which will be visualized using the interactive tree viewer embedded within the online article. Using the viewer it will be possible to zoom into certain tree areas, change the tree layout, search within the tree, and collapse/expand tree nodes and branches. Submitted tree files will also be available for downloading from your online article on ScienceDirect. Each tree must be contained in an individual data file before being uploaded separately to the online submission system, via the 'phylogenetic tree data' submission category. Newick files must have the extension .new or .nwk (note that a semicolon is needed to end the tree). Please do not enclose comments in Newick files and also delete any artificial line breaks within the tree data because these will stop the tree from showing. For NeXML, the file extension should be .xml. Please do not enclose comments in the file. Tree data submitted with other file extensions will not be processed. Please make sure that you validate your Newick/NeXML files prior to submission. For more information please see http://www.elsevier.com/phylogenetictrees.

\bigskip \noindent \textbf{\emph{Nature} (weak policy$^\dagger$)}

\noindent
Therefore, a condition of publication in a Nature journal is that \emph{authors are required to make materials, data and associated protocols promptly available to readers} without undue qualifications.

The preferred way to share large data sets is via public repositories.

\noindent 
NB: $^\dagger$Data-sharing policy in place at the time of study publication, which has since become \emph{JDAP membership.}

\bigskip \noindent \textbf{\emph{New Phytologist} (weak policy)}

\bigskip \noindent \textbf{\emph{Perspectives in Plant Ecology, Evolution and Systematics} (no policy)}

\noindent
Elsevier \emph{accepts electronic supplementary material to support and enhance} your scientific research. Supplementary files offer the author additional possibilities to publish supporting applications, high-resolution images, background datasets, sound clips and more.

You can \emph{enrich your online articles} by providing phylogenetic tree data files (optional) in Newick or NeXML format, which will be visualized using the interactive tree viewer embedded within the online article.

Electronic archiving of supplementary data enables readers to replicate, verify and build upon the conclusions published in your paper. We\emph{ recommend that data should be deposited} in the data library PANGAEA (http://www.pangaea.de).

\bigskip \noindent \textbf{\emph{Plant Systematics and Evolution} (weak policy)}

\noindent
Data matrices \emph{including sequence alignments must be made available to the public.} There must be a sentence included in the Materials and methods section that such information is available from the corresponding author. \emph{“DNA or proteine sequences must be deposited in public data bases} (GenBank, EMBL, etc.) before the revised version is sent to the editor.“

\bigskip \noindent \textbf{\emph{PLOS ONE} (weak policy$^\dagger$)}

\noindent
Manuscripts reporting paleontology and archaeology research must include descriptions of methods and specimens in sufficient detail to allow the work to be reproduced. \emph{Data sets supporting statistical and phylogenetic analyses should be provided, preferably in a format that allows easy re-use.}

\emph{Specimen numbers and complete repository information,} including museum name and geographic location, \emph{are required for publication.} Locality information should be provided in the manuscript as legally allowable, or a statement should be included giving details of the availability of such information to qualified researchers.

\emph{Methods sections of papers with data that should be deposited} in a publicly available database \emph{should specify where the data have been deposited} and provide the relevant accession numbers and version numbers, if appropriate. Accession numbers should be provided in parentheses after the entity on first use. If the accession numbers have not yet been obtained at the time of submission, please state that they will be provided during review. They must be provided prior to publication.

\noindent 
NB: $^\dagger$Data-sharing policy in place at the time of study publication, which has since become \emph{JDAP membership.}

\bigskip \noindent \textbf{\emph{PLOS Biology} (weak policy$^\dagger$)}

\noindent
The results section should provide details of all of the experiments that are required to support the conclusions of the paper.

\emph{Large datasets, including raw data, should be submitted as supplemental files;} these are published online alongside the accepted article.

\noindent 
NB: $^\dagger$Data-sharing policy in place at the time of study publication, which has since become \emph{JDAP membership.}

\bigskip \noindent \textbf{\emph{Proceedings of the National Academy of Sciences} (weak policy)}

\noindent
To allow others to replicate and build on work published in PNAS, \emph{authors must make materials, data, and associated protocols available to readers}. Authors must disclose upon submission of the manuscript any restrictions on the availability of materials or information. Data not shown and personal communications cannot be used to support claims in the work. Authors are encouraged to use SI to show all neces- sary data. Authors are encouraged to deposit as much of their data as possible in publicly accessible databases.

Before publication, \emph{authors must deposit large datasets (including microarray data, protein or nucleic acid sequences, and atomic coordinates for macromolecular structures) in an approved database} and provide an accession number for inclusion in the published paper. \emph{When no public repository exists, authors must provide the data as SI online} or, in special circumstances when this is not possible, on the author’s institutional Web site, provided that a copy of the data is provided to PNAS.

\emph{Authors must deposit data in a publicly available database} such as GenBank, EMBL, DNA Data Bank of Japan, UniProtKB/Swiss-Prot, or PRIDE.

\bigskip \noindent \textbf{\emph{Proceedings of the Royal Academy B: Biological Sciences} (strong policy)}

\noindent
To allow others to verify and build on the work published in Royal Society journals \emph{it is a condition of publication that authors make available the data and research materials supporting the results in the article,} as detailed in our Publishing policies; with which our authors are asked to comply.

\emph{Datasets should be deposited in an appropriate, recognized repository and the associated accession number, link or DOI to the datasets must be included in the methods section of the article.} Reference(s) to datasets should also be included in the reference list of the article with DOIs (where available). Where no discipline-specific data repository exists authors should deposit their datasets in a general repository such as Dryad (http://datadryad.org/)

Where possible any other relevant research materials (such as statistical tools, protocols, software etc) should also be made available and details of how they may be obtained should be included in the data accessibility section of the article.

To ensure archived data from Proceedings B articles are available to readers, \emph{authors should include a ‘data accessibility’ section} immediately after the acknowledgements. This should list the database and accession number for all data from the article that has been made publicly available.

\bigskip \noindent \textbf{\emph{Science} (strong policy)}

\noindent
\emph{All data necessary to understand, assess, and extend the conclusions of the manuscript must be available to any reader of Science.} All computer codes involved in the creation or analysis of data must also be available to any reader of Science. \emph{After publication, all reasonable requests for data and materials must be fulfilled.}

Science supports the efforts of databases that aggregate published data for the use of the scientific community. Therefore, \emph{appropriate data sets (including microarray data, protein or DNA sequences, atomic coordinates or electron microscopy maps for macromolecular structures, and climate data) must be deposited in an approved database, and an accession number or a specific access address must be included in the published paper}. We encourage compliance with MIBBI guidelines (Minimum Information for Biological and Biomedical Investigations).

\emph{Large data sets with no appropriate approved repository must be housed as supplementary materials} at Science, or only when this is not possible, on an archived institutional Web site, provided a copy of the data is held in escrow at Science to ensure availability to readers.

\bigskip \noindent \textbf{\emph{Systematic Biology} (strong policy)}

\noindent
\emph{All nucleotide sequence data and alignments must be submitted to GenBank or EMBL} before the paper can be published. In addition, \emph{all data matrices and resulting trees must be submitted to TreeBASE.} GenBank and TreeBASE \emph{reference numbers should be provided in the final version of the paper.}

\bigskip \noindent \textbf{\emph{Systematic Entomology} (no policy)}

\noindent
Paper submissions will be accepted exceptionally, although any relevant data matrices should be electronic. 

\bigskip \noindent \textbf{\emph{Trends in Microbiology} (no policy)}

\noindent
A Trends in Microbiology Reviews \emph{must not include unpublished data, simulations or meta-analyses,} or propose a new formal mathematical model.

A Trends in Microbiology Opinion \emph{must not include unpublished data, simulations or meta-analyses,} or propose a new formal mathematical model.

Science \& Society articles \emph{should not include unpublished data, simulations or meta-analyses.}

Letters should \emph{not be used as an opportunity} to promote your own work, \emph{to introduce new data,} to provide an 'update' on a recent review, or to highlight perceived omissions in our articles.

\bigskip \noindent \textbf{\emph{Virus Research} (no policy)}

\noindent
Elsevier \emph{encourages authors to connect articles with external databases,} giving their readers one-click access to relevant databases that help to build a better understanding of the described research. 

Elsevier \emph{accepts electronic supplementary material to support and enhance your scientific research.} Supplementary files offer the author additional possibilities to publish supporting applications, high-resolution images, background datasets, sound clips and more. Supplementary files supplied will be published online alongside the electronic version of your article in Elsevier Web products, including ScienceDirec

\bigskip \noindent \textbf{\emph{Zoological Journal of the Linnean Society} (no policy)}

\noindent
\emph{Avoid elaborate tables of original or derived data}, long lists of species, etc.; \emph{if} such data are \emph{absolutely essential,} consider including them as appendices or as online-only supplementary material. 

\end{document}